\documentclass[final,3p]{elsarticle}

\usepackage{hyperref}

\usepackage[usenames,dvipsnames]{color}
\usepackage[english]{babel}
\usepackage{graphicx}
\usepackage{amssymb}
\usepackage{amsmath}
\usepackage{amstext}
\usepackage{bm}

\begin{document}

\def\be{\begin{equation}}
\def\ee{\end{equation}}

\def\St{\text{St}}
\def\ETh{E_{\text{Th}}}
\def\mls{\Delta}
\def\br{{\bf r}}
\def\bp{{\bf p}}
\def\bq{{\bf q}}
\def\bn{{\bf n}}
\def\Lam{\Lambda}
\def\vp{\varphi}
\def\eps{\varepsilon}
\def\gammaRPA{\gamma_0^\text{RPA}}
\def\pf{p_{F}}
\def\Re{\mathop{\rm Re}}
\def\Im{\mathop{\rm Im}}
\def\sign{\mathop{\rm sign}}
\def\ve{\varepsilon}
\def\FB{\text{FB}}
\def\RA{\text{RA}}
\def\AR{\text{AR}}
\def\RR{\text{RR}}
\def\AA{\text{AA}}
\def\str{\mathop{\rm str}}
\def\tr{\mathop{\rm tr}}
\def\B{\text{B}}
\def\F{\text{F}}

\newcommand{\corr}[1]{\langle #1\rangle}
\newcommand{\ccorr}[1]{\langle\langle #1\rangle\rangle}
\newcommand{\com}[1]{\textbf{\color{red}(#1)}}

\newcommand{\tit}[1]{\emph{#1}}

\def\frs{F}
\def\ns{N_s}
\def\EFGR{\varepsilon_\text{FGR}}
\def\EMBL{\varepsilon_\text{MBL}}
\def\FF{{\cal F}}
\def\BB{{\cal B}}

\begin{frontmatter}

\title{Energy relaxation rate and its mesoscopic fluctuations in quantum dots%
}

\author[mit]{Vladyslav A. Kozii}
\author[skoltech,itp,mipt]{Mikhail A. Skvortsov}

\address[mit]{Department of Physics, Massachusetts Institute of Technology, Cambridge,
Massachusetts 02139, USA}
\address[skoltech]{Skolkovo Institute of Science and Technology, Skolkovo 143026, Russia}
\address[itp]{L. D. Landau Institute for Theoretical Physics,
Chernogolovka 142432, Russia}
\address[mipt]{Moscow Institute of Physics and Technology, Dolgoprudny 141700, Russia}

\date{\today}

\begin{abstract}
We analyze the applicability of the Fermi-golden-rule description of quasiparticle relaxation in a closed diffusive quantum dot with electron-electron interaction.
Assuming that single-particle levels are already resolved but the initial stage of quasiparticle disintegration can still be described by a simple exponential decay,
we calculate the average inelastic energy relaxation rate of single-particle excitations and its mesoscopic fluctuations. The smallness of mesoscopic fluctuations can then be used as a criterion for the validity of the Fermi-golden-rule description. Technically, we implement the real-space Keldysh diagram technique, handling correlations in the quasi-discrete spectrum non-perturbatively by means of the non-linear supersymmetric sigma model.
The unitary symmetry class is considered for simplicity. Our approach is complementary to the lattice-model analysis of Fock space: thought we are not able to describe many-body localization, we derive the exact lowest-order expression for mesoscopic fluctuations of the relaxation rate, making no assumptions on the matrix elements of the interaction.
It is shown that for the quasiparticle with the energy $\eps$ on top of the thermal state with the temperature $T$, fluctuations of its energy width become large and the Fermi-golden-rule description breaks down at $\max\{\eps,T\}\sim\Delta\sqrt{g}$, where $\Delta$ is the mean level spacing in the quantum dot, and $g$ is its dimensionless conductance.
\end{abstract}

\end{frontmatter}

\tableofcontents

\section{Introduction}

Understanding the structure of wave functions in Fock space is a hot topic in many-body physics, intimately related to the concepts of ergodicity and thermalization in complex interacting systems.
The natural way to think of the interaction is as a cause of transitions between single-particle states, which are no longer eigenstates of a quantum system in the presence of interaction.
Singe-particle excitations decay into three-particle states (two electrons and one hole), which then further decay into five-particle states, etc. The simplest way to characterize the process of quantum-mechanical spreading of an excitation in Fock space is to study its energy relaxation rate (inverse lifetime).

The study of the inelastic relaxation in a confined geometry has been pioneered by Sivan, Imry and Aronov (SIA) \cite{SIA}, who calculated the quasiparticle lifetime in a diffusive quantum dot with chaotic electron dynamics.
Working within the conventional Fermi golden rule (FGR) picture, they calculated the relaxation rate of an excitation with the energy $\ve$, induced by the screened Coulomb interaction with the small momentum transfer:
\be
  \gamma_0(\ve)
  \sim
  \lambda^2 \Delta (\ve/\ETh)^2 ,
\label{SIA}
\ee
where $\Delta$ is the mean single-particle level spacing in the dot, and $\ETh= D/L^2$ is the Thouless energy determined by the inverse time of electron diffusion across the system ($D$ is the diffusion coefficient, and $L$ is the typical size of the quantum dot).
For generality, in Eq.~(\ref{SIA}) we introduced the dimensionless interaction strength $\lambda$, taking its maximal value, $\lambda=1$, for the screened Coulomb potential.
The result (\ref{SIA}) was derived in the hot-electron regime (negligible temperature, $T\ll\ve$) under the assumption of the zero-dimensional geometry, $\ve \ll \ETh$.
It provides the diffusive contribution to the relaxation rate originating from the processes with momentum transfer of the order of the inverse system size, $1/L$. The total relaxation rate is then the sum of $\gamma_0(\ve)$ given by Eq.~(\ref{SIA}) and the Fermi-liquid contribution due to the processes with large-momentum transfer, $\gamma_*(\ve)\sim \eps^2/E_F$ ($E_F$ is the Fermi energy) \cite{FL}. The latter can be neglected for sufficiently large quantum dots, $L\gg k_Fl^2$ ($k_F$ is the Fermi momentum, and $l$ is the mean free path) \cite{SIA,Blanter}, that will be assumed thereafter.

Remarkably, Eq.~(\ref{SIA}) derived in the zero-dimensional limit, $\ve \ll \ETh$, shows that in this regime the single-particle spectrum is well resolved, $\gamma_0(\ve)\ll\Delta$. Though this conclusion is in a good agreement with the experimental results on tunneling spectroscopy of a disordered quantum dot~\cite{SMM}, it raises the question of consistency of the derivation.
Indeed, the FGR can be safely applied only in the case of continuous spectrum, while the result (\ref{SIA}) implies that it is not. This subtle point was recognized already by SIA, who argued that their approach might be correct as summation over many final states is performed.
This idea was elaborated in a seminal paper by Altshuler, Gefen, Kamenev, and Levitov (AGKL) \cite{AGKL}, who emphasized that it is the spectrum of final states that should be continuous for the FGR picture to be applicable. In the problem of the hot-electron decay, final states are three-particle states with two electrons and one hole, and the corresponding mean level spacing can be estimated as
\be
  \Delta_3(\ve) \sim \Delta^3/\ve^2
  .
\label{Delta3}
\ee
Comparing $\gamma_0(\eps)$ and $\Delta_3(\eps)$, AGKL came to the conclusion that the FGR description should be valid for sufficiently large energies, $\ve>\EFGR$, where
\be
\EFGR=\Delta\sqrt g,
\label{eps*}
\ee
and $g=\ETh/\lambda\Delta\gg1$ sets the scale, $\Delta/g$, of the interaction matrix elements (for the screened Coulomb interaction with $\lambda=1$, $g$ coincides with the dimensionless conductance of the dot).

In order to study the spreading of a single-particle excitation over the space of many-particle states beyond the FGR approximation,
AGKL proposed a remarkable mapping of the initial problem to a tight-binding Anderson model on a hierarchical lattice, treating each site as a basis vector in the many-particle Fock space. Motivated by the observation that the structure of this lattice for the quantum dot problem locally looks like a tree, AGKL approximated it by the Bethe lattice with a large branching number. Then using the known solution for the Anderson model on the Bethe lattice \cite{Cayley}, AGKL predicted the localization transition in Fock space of an interacting quantum dot at the energy $\EMBL\sim\Delta\sqrt{g/\ln g}$, which appeared to be parametrically smaller than the FGR breaking energy scale $\EFGR$. Thus, in the AGKL picture of many-body localization one should distinguish between three regimes \cite{AGKL,MF97}.
In the localized regime realized at $\ve<\EMBL$, the Slater determinants of one-particle states are very close to the exact many-body states, and the width of quasiparticle states is exactly zero.
In the intermediate region, $\EMBL<\eps<\EFGR$, quasiparticle states are delocalized, but they are strongly fractal and non-ergodic, with the spectral weight given by a number of slightly broadened lines.
The spectral weight acquires a Lorentzian form with a well defined width $\gamma$ (corresponding to a simple exponential decay $e^{-\gamma t}$ in the time domain) only in the regime $\eps\gg\EFGR$, where the FGR description finally sets in.

The AGKL paper has triggered a boost of activity in the field of many-body localization. The suggested mapping onto the lattice model has been proved to be extremely fruitful: instead of studying an interacting problem one now can deal with a non-interacting quantum mechanics, but on a very complicated lattice with an exponentially large number of sites.
(Though all states in a finite system are localized by definition, even an extremely weak coupling to the external reservoir providing a level width larger than the exponentially small distance between the many-body states, which can be estimated as $\exp(-\alpha\sqrt{\eps/\Delta})$ with $\alpha\sim1$ \cite{Silv,Bethe1936,BohrMottelson},
renders the spectrum of delocalized states effectively continuous.)
Since the complexity of the problem is encoded in the lattice topology, the crucial point is
to identify the relevant features responsible for many-body localization.

Quite soon it was recognized that the Bethe lattice with a constant branching number is an oversimplified model of Fock space.
Firstly, the coordination number of the lattice decreases with the number of generations \cite{Silv,Jacquod97}. Secondly, the actual lattice is not a tree, and the presence of loops essentially modifies combinatorics of the perturbative expansion in the localized region \cite{Silv}, increasing the AGKL estimate for $\EMBL$.
Numerical studies \cite{LTB98,Kamenev2002} demonstrated gradual delocalization with the growth of the quasiparticle energy.
Anyway, despite the lack of a rigorous theory of many-body localization in a quantum dot, the general understanding achieved in the beginning of 2000s was that in finite systems one should expect a localization/delocalization crossover, though its position was not firmly established.

Later, the concept of many-body localization developed in the quantum dot problem
was applied to extended systems of interacting electrons with spatially localized single-particle states \cite{BAA,GMP}, where the transition between the localized and delocalized many-body phases occurs at a finite temperature.
The related issues of ergodicity and thermalization are currently being actively investigated (for a review, see Ref.~\cite{Polkovnikov}).

Very recently, many-body localization in a quantum dot was reconsidered by Gornyi, Mirlin, and Polyakov \cite{Mirlin2015}. Working in the framework of the lattice model, they found an additional factorial contribution in the perturbative series in the number of involved generations, which renders coupling to distant generations less efficient, thus acting in favor of localization.
For the hot-electron decay problem, this leads to the estimate for the threshold energy $\EMBL\sim g\mls/\ln g$, which is much larger than the original AGKL estimate.
For a thermal many-body state with the temperature $T$ and the total energy ${\cal E}\sim T^2/\mls$, the FGR temperature $T_\text{FGR}\sim \mls\sqrt{g}$ was shown to be logarithmically larger than the localization transition point, $T_\text{MBL}\sim\mls\sqrt{g/\ln g}$. This result formally  coincides with the AGKL \emph{energy}\ threshold and is in full agreement with Ref.~\cite{BAA}.

Since, contrary to AGKL, Refs.~\cite{Silv,Mirlin2015} place the many-body localization threshold $\EMBL$ for the hot-electron problem above the FGR-breaking scale $\EFGR$,
one can ask how to reconcile the FGR description with many-body localization. This question was answered by Silvestrov \cite{Silvestrov2001} who studied the temporal decay of a quasiparticle with $\eps\gg\EFGR$. He showed that the FGR relaxation rate $\gamma_0(\eps)$ describes the initial stage of exponential relaxation, that slows down at larger times due to smaller relaxation rate of descendant states, and eventually due to many-body localization.

In this paper, we address the initial temporal stage of energy relaxation in a quantum dot, when quantum many-body localization effects are not yet visible. Assuming the FGR approach is applicable, we calculate mesoscopic fluctuations of the energy relaxation rate. The smallness of fluctuations compared to the average relaxation rate provides an \emph{a posteriori}\ condition for the FGR applicability. In our analysis, we do not use the lattice model of Fock space and work in terms of Keldysh diagram technique in real space, where non-perturbative disorder averaging is performed by means of the non-linear supersymmetric sigma model. Such an approach allows us to derive the expression for the variance of the relaxation rate without any simplifications concerning the nature of the electron-electron interaction.

The paper is organized as follows. In Sec.~\ref{S:Model} we introduce the model and summarize the results. Section~\ref{S:General} describes the Keldysh kinetic approach
along with its modification simplifying the study of weak nonequilibrium.
In Sec.~\ref{S:Average} we rederive the SIA result and generalize it to the case of an arbitrary temperature and interaction radius. Section~\ref{S:MF} is devoted to the discussion of the general strategy for the calculation of mesoscopic fluctuations of the energy relaxation rate in terms of the exact (disorder-dependent) electron Green functions. The product of several Green functions is averaged non-perturbatively in Sec.~\ref{S:nonpert}. The final step of the calculation of mesoscopic fluctuations is performed, for not very short-range interaction, in Sec.~\ref{S:final} and, in the general case, in Sec.~\ref{S:mesofluct-rs}. Our results are summarized in Sec.~\ref{Discussion}. Numerous technical details are relegated to several Appendices.

We use the system of units with $\hbar=k_B=1$.

\section{Model description and results}
\label{S:Model}

\subsection{Model}

We consider an isolated chaotic quantum dot with the broken time-reversal symmetry (unitary symmetry class).
Impurity scattering in the dot is supposed to be strong enough to completely
randomize the electron trajectories establishing the diffusive regime with
$l\ll L$, where $l$ is the elastic mean free path,
and $L$ is the characteristic size of the dot.
For generality, we consider the case of an arbitrary spin degeneracy, $\ns$, which plays the role of the number of independent fermionic flavors (the case $\ns=1$ corresponds to completely spin-polarized electrons).

We will discuss two models of electron-electron interactions: (i) the long-range Coulomb interaction with a small gas parameter $r_s \equiv e^2/\epsilon v_F \ll 1$ (where $\epsilon$ is the dielectric constant of the medium, and $v_F$ is the Fermi velocity), and (ii) an arbitrary weak interaction. In the case of the Coulomb interaction, its screening should be taken into account, whereas for a weak interaction this procedure is unnecessary.
In both cases, the statically screened interaction in the real space can be written in the form
\be
\label{V(r)}
  V(\br)
  =
  \frac{\lambda}{\ns\nu} \delta_\kappa(\br) ,
\ee
where $\lambda$ is the dimensionless interaction strength,
and $\delta_\kappa(\br)$ is the delta function smeared on the scale of the interaction radius $1/\kappa$
[the latter is assumed to be much smaller than the size of the quantum dot,
justifying the notion of the smeared delta function in Eq.~(\ref{V(r)})].
For the long-range Coulomb interaction, $\lambda=1$ and $1/\kappa$ is the Thomas-Fermi screening length,
see Eq.~(\ref{kappa}).
Besides the dimensionless strength $\lambda$, the only characteristic of the potential (\ref{V(r)}) that will be important in the following is the ratio, $\frs$, of the Hartree and Fock diagrams, which is a measure of the interaction range \cite{Altshulerbook,Akkermans}. In the three- and two-dimensional cases it is given by (see \ref{S:Non-RPA} for details)
\be
\label{f-gen}
  \frs
  =
  \int_0^{1} v_\kappa(2p_Fx) \varphi(x) \, dx ,
\qquad
  \varphi(x)
  =
  \begin{cases}
    2 x, & \text{in 3D}, \\
    (2/\pi) (1-x^2)^{-1/2}, & \text{in 2D}, \\
  \end{cases}
\ee
where $v_\kappa(\bq)$ is the Fourier transform of $\delta_\kappa(\br)$ in Eq.~(\ref{V(r)}).
The limiting cases of $\frs=0$ and $\frs=1$ correspond to long-range ($\kappa\ll k_F$) and point-like screened interactions, respectively.

We will assume that interaction can be taken into account perturbatively, that can be justified in the two partially overlapping limits:

\be
\label{lambda-f}
  \text{$\lambda\ll1$ or $\frs\ll1$} ,
\ee
i.e., in the limit of weak interaction or for sufficiently large interaction radius ($\kappa\ll k_F$).
The two models of electron-electron interaction discussed above conform to the condition (\ref{lambda-f}):
For the Coulomb interaction with $\lambda=1$, its applicability is provided by the smallness of the gas parameter $r_s$ since $\frs\sim r_s\ln 1/r_s$ [see Eq.~(\ref{f-Coulomb})]; for weak interaction with an arbitrary radius $1/\kappa$, it is justified as long as $\lambda\ll1$.

\subsection{Average energy relaxation rate}

We start with generalizing the SIA result (\ref{SIA}) to the case of a finite temperature $T$, arbitrary spin degeneracy $\ns$ and arbitrary interaction radius measured by the parameter $\frs$ [assuming that the restriction (\ref{lambda-f}) holds].
The temperature and excitation energy are supposed to be smaller than the Thouless energy, $\max \{ \ve, T \}\ll \ETh$.
The average energy relaxation rate calculated in the second order in the statically screened interaction (\ref{V(r)}) is given by
\be
  \gamma_0(\ve,T)
  =
  \frac{\lambda^2c(\ns,\frs)}{\pi}
  \frac{\Delta}{E_2^2} (\ve^2+\pi^2 T^2) ,
\label{SIAT}
\ee
where $\Delta=(\nu V)^{-1}$ is the mean level spacing in the dot
($\nu$ is the single-particle density of states at the Fermi level per one spin projection, and $V$ is volume of the dot), and the energy scale $E_2$ is determined by the diffusion inside the dot:
\be
\label{En}
  E_n\equiv\biggl[\mathop{{\sum}'}_m \frac{1}{(Dq^2_m)^n}\biggr]^{-1/n} \sim \ETh
  .
\ee
Here $D$ is the diffusion coefficient, and $q^2_m$ are non-zero eigenvalues
of the Laplace operator in the dot, $- \nabla^2 \psi_m(\br) = q_m^2 \psi_m(\br)$,
with the von Neumann boundary conditions.
The magnitude of $E_n$ defined by Eq.~(\ref{En}) is set by the Thouless energy,
$\ETh=D/L^2$, where $L$ is the typical size of the quantum dot.

The factor $c(\ns,\frs)$ in Eq.~(\ref{SIAT}) takes into account spin degeneracy and the spatial structure of the interaction potential (\ref{V(r)}):
\be
\label{c-res}
  c(\ns,\frs)
  =
  \frac{1}{\ns^2} \left(\ns - 2\frs + \ns \frs^2 \right) .
\ee
The function $c(\ns,\frs)$ has an important property $c(1,1)=0$, which ensures vanishing of interaction effects for polarized ($\ns=1$) fermions with a contact interaction,
as a consequence of the Pauli exclusion principle. It should be noted
that the anticipated cancellation of $c(1,1)$ takes place only if the Hartree-type diagrams yielding the terms with $\frs$ in Eq.~(\ref{c-res}) are taken into account (see Sec.~\ref{SS:rs}). This issue is often overlooked in literature, leading to a variety of claimed prefactors in Eq.~(\ref{SIA}).

\subsection{Mesoscopic fluctuations of the energy relaxation rate}

In the FGR description, the relaxation rate $\gamma_i$ of a given single-particle state $|i\rangle$ is a function of its energy only, implying that the states close in energy should have nearly the same inelastic width.
The validity of this approximation can be verified \emph{a posteriori}\
by studying mesoscopic, i.~e., level-to-level fluctuations of $\gamma_i$.
We calculate them under the same assumptions of the zero-dimensional geometry,
$\max \{ \ve, T \}\ll \ETh$, used for the determination of the average $\gamma_0(\eps,T)$.

We describe the relaxation rate in the framework of the quantum kinetic equation for electrons in dirty metals~\cite{RS}. This approach differs significantly from the lattice model proposed by AGKL \cite{AGKL} and extensively used in subsequent publications \cite{MF97,Silv,Jacquod97,LTB98,Mirlin2015}.
The method of kinetic equation is not suitable for studying many-body localization, but is sufficiently simple to unveil the limits of applicability of the FGR description.
Nevertheless, the kinetic approach, working well for interacting systems with continuous spectrum, meets significant difficulties in the considered region $\max\{\ve,T\}\ll\ETh$, where individual energy levels are well-resolved, $\gamma_{0}(\ve,T)\ll \Delta$.
Discreteness of energy spectrum requires non-perturbative averaging over disorder, which is performed by means of the non-linear supersymmetric sigma model \cite{Efetov-book}.

The main idea behind our calculations is the following. In the zeroth approximation (which is equivalent to the FGR), energy levels acquire an inelastic width $\gamma_0(\ve,T)$. The inverse of this quantity yields the temporal scale
at which single-particle coherence is maintained.
This corresponds to the appearance of the `mass' of the order $\gamma_{0}(\ve,T)$ for the zero-dimensional diffusons (diffusons with zero momentum).
At the next stage, when loop corrections to the FGR result are considered, this `mass' will regularize the otherwise divergent contributions, producing $\gamma_0(\ve,T)$ in the denominator, precisely in the way it occurs in the case of the non-interacting quantum dot in an external field \cite{dyn-loc,Kubo,Kubo4} and in the high-temperature phase in the problem of many-body localization \cite{BAA}. Thus, with decreasing the energy of excitations and/or temperature, the loop corrections to the quasiclassical rate increase, and at some energy scale they become comparable, indicating the breakdown of the FGR description.

Fluctuations of the energy relaxation rate $\gamma(\ve,T)$ around its FGR average value $\gamma_0(\ve,T)$ given by Eq.~(\ref{SIAT}) are characterized by the irreducible average:
\be
\label{corr-irr-def}
  \corr{\gamma^2(\ve,T)}=\gamma_0^2(\ve,T) + \ccorr{\gamma^2(\ve,T)} .
\ee
We calculate the leading contribution to $\ccorr{\gamma^2(\ve, T)}$ in the delocalized regime and obtain
\be
  \ccorr{\gamma^2(\ve,T)}
  =
  \frac{\lambda^4\Delta^5}{4\pi^3}
  \left[\frac{c_2(\ns,\frs)}{E_2^4}+\frac{c_4(\ns,\frs)}{E_4^4}\right]
  \int d\ve_1 d\omega_1
  \frac{(\FF_{\ve_1}-\FF_{\ve_1-\omega_1})^2(\BB_{\omega_1}+\FF_{\ve-\omega_1})^2}
  {\gamma_0(\ve-\omega_1,T)+\gamma_0(\ve_1,T)+\gamma_0(\ve_1-\omega_1,T)}
  ,
\label{subfinal}
\ee
where
\be
\label{FandB}
  \FF_{\ve}=\tanh \frac{\ve}{2T},
  \qquad
  \BB_{\omega}=\coth\frac{\omega}{2T}
\ee
are equilibrium fermionic and bosonic distribution functions, respectively, while the energies $E_2$ and $E_4$ (both of the order of $\ETh$) are defined in Eq.~(\ref{En}). The functions $c_2(\ns,\frs)$ and $c_4(\ns,\frs)$ in Eq.~(\ref{subfinal}) are given by
\begin{subequations}
\label{c2c4}
\begin{gather}
\label{c2}
  c_2(\ns,\frs)
  =
  \frac{1}{\ns^4}
  \left[
    (3\ns^2+1) - 16\ns \frs + 2(5\ns^2+7) \frs^2
  - 16\ns \frs^3 + (3\ns^2+1) \frs^4
  \right]
,
\\
\label{c4}
  c_4(\ns,\frs)
  =
  \frac{1}{\ns^4}
  \left[
    2\ns^2 - 8\ns \frs + 4(\ns^2+2) \frs^2
  - 8\ns \frs^3 + 2\ns^2 \frs^4
  \right] .
\end{gather}
\end{subequations}
They obey an important relation $c_2(1,1)=c_4(1,1)=0$, which guarantees vanishing of mesoscopic fluctuations [as well as the inelastic width itself, see Eq.~(\ref{c-res})] for spin-polarized fermions with a contact interaction.

\begin{figure}
\centerline{\includegraphics[width=.55\textwidth]{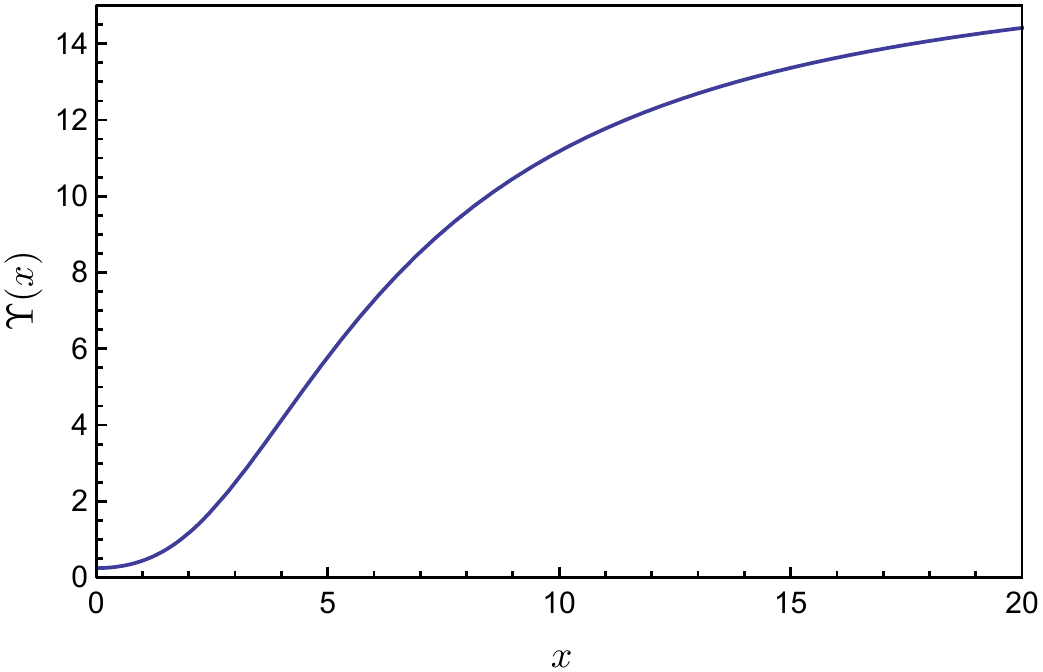}}
\caption{The function $\Upsilon(x)$ [Eq.~(\ref{Upsilon-def})], which determines the energy and temperature dependence of $\ccorr{\gamma^2(\ve,T)}$, see Eq.~(\ref{gamma-Upsilon}).}
\label{F:Upsilon}
\end{figure}

To single out the energy and temperature dependence of Eq.~(\ref{subfinal}), we rewrite it in the form
\be
\label{gamma-Upsilon}
  \ccorr{\gamma^2(\ve,T)}
  =
  \frac{\lambda^2 \Delta^4}{4\pi^2} \frac{E_2^2}{c(\ns,\frs)} \left[\frac{c_2(\ns,\frs)}{E_2^4}+\frac{c_4(\ns,\frs)}{E_4^4}\right]
  \Upsilon \left( \frac{\ve}{2T} \right),
\ee
where the dimensionless function $\Upsilon(x)$ plotted in Fig.~\ref{F:Upsilon} is defined as
\be
\label{Upsilon-def}
  \Upsilon(x)
  =
  \int dy \, dz \,
  \frac{[\tanh(y)-\tanh (y-z)]^2[\coth(z)-\tanh(z-x)]^2}{(x-z)^2+y^2+(y-z)^2 + 3\pi^2/4}
  =
  \begin{cases}
    0.248 , & x = 0 ; \\
    16.95 , & x = \infty .
  \end{cases}
\ee
Hence, the magnitude of mesoscopic fluctuations of the relaxation rate can be roughly estimated as
\be
  \ccorr{\gamma^2(\ve,T)}
  \sim
  \lambda^2
  \frac {\Delta^4}{\ETh^2} .
\label{qualresult1}
\ee
Note however that, due to a huge variation of the function $\Upsilon(x)$, fluctuations at the Fermi energy ($\eps\ll T$) are nearly 100 times smaller than the naive estimate (\ref{qualresult1}).

Expression (\ref{subfinal}) for mesoscopic fluctuations of the energy relaxation rate in the FGR regime is the main result of our work.
The relative strength of mesoscopic fluctuations can be estimated as
\be
  \frac{\ccorr{ \gamma^2(\ve,T)}}{\gamma_{0}^2(\ve,T)}
  \sim
  \frac{\Delta_3(\max\{\ve,T\})}{\gamma_{0}(\ve,T)}
  \sim
  \left( \frac{\EFGR}{\max\{ \ve, T\}} \right)^4 ,
\label{qualresult2}
\ee
where $\Delta_3(\ve)$ is the mean three-particle level spacing defined in Eq.~(\ref{Delta3}). The ratio (\ref{qualresult2}) becomes of the order of unity as $\max\{\ve,T\}$ approaches the FGR-breaking scale $\EFGR = \Delta\sqrt{g}$. Hence for the validity of the FGR description of the initial stage of quasiparticle disintegration, the temperature $T$ of electrons in the dot and the excitation energy $\ve$ at which the decay rate is studied play nearly the same role. They will act differently at larger time scales, when many-body localization may have enough time to show up \cite{Mirlin2015,Silvestrov2001}.
Physically, at $\max\{\ve,T\}\gg\EFGR$ the state has many available routes to decay into three-particle excitations. Fluctuations of the number of such routes are small and are determined by Eq.~(\ref{qualresult2}).

\section{General expression for the energy relaxation rate}
\label{S:General}

\subsection{Keldysh technique}
\label{SS:Keldysh}

In order to express the inelastic energy relaxation rate, we use the Keldysh technique \cite{Keldysh} in the representation of the functional integral \cite{KA99,KL09}. For an interacting system, the Keldysh action
is a functional of the fermionic Grassmann fields $\psi_\sigma(\br,t)$
($\sigma=1,\dots,\ns$ counts spin projections)
and the bosonic plasmon field $\phi(\br,t)$:
\be
\label{action1}
  S =
  \int dt\left\{\sum_{\sigma=1}^{\ns} \int d\br \, \psi^+_{\sigma}(\br,t) \left( \hat G_0^{-1} + \phi_a(\br,t) \gamma^a \right) \psi_{\sigma}(\br,t)
+ \int(d\bq) V_0^{-1}(\bq) \, \phi^T(\bq,t)\gamma_2\phi(-\bq,t)\right\}.
\ee
The fields $\psi_\sigma(\br,t)$ and $\phi(\br,t)$ are two-component vectors in the Keldysh space.
In the Keldysh-rotated basis \cite{KA99,KL09},
\be
\hat G_0^{-1}=\left(i\frac{\partial}{\partial t}+\frac{\nabla^2}{2m}-U_{\text{dis}}(\br)\right)\gamma_1\equiv G_0^{-1}\gamma_1,
\qquad
\gamma^1=\begin{pmatrix} 1&0\\0&1 \end{pmatrix},
\qquad
\gamma^2=\begin{pmatrix} 0&1\\1&0 \end{pmatrix},
\ee
where $U_{\text{dis}}(\br)$ is the disorder potential
with the correlation function $\corr{U_{\text{dis}}(\br)U_{\text{dis}}(\br')} = \delta(\br-\br')/(2\pi\nu\tau)$
($\tau$~is the elastic mean free time, and $\nu$ is the density of states per one spin projection
at the Fermi energy). We assume no spin-dependent interactions so that all matrices
act as a unit matrix in the spin space.
The second term in Eq.~(\ref{action1}) is the action of the plasmon field,
with $V_0(\bq)$ being the bare interaction potential,
and $(d\bq)\equiv d^dq/(2\pi)^d$ is the momentum integration measure in $d$ dimensions.

The Green functions are defined as ($\xi_i$ denotes the pair $\br_i,t_i$):
\begin{gather}
\hat G(\xi_1,\xi_2)=-i\langle\psi(\xi_1)\psi^+(\xi_2)\rangle
=-i\int D\psi^{*}D\psi D\phi \, e^{iS} \, \psi(\xi_1)\psi^+(\xi_2),
\\
\hat V(\xi_1,\xi_2)=-2i\langle\phi(\xi_1)\phi^T(\xi_2)\rangle
=-2i\int D\psi^{*}D\psi D\phi \, e^{iS} \, \phi(\xi_1)\phi^T(\xi_2) .
\label{propV}
\end{gather}
The electron Green function has the triangular structure in the Keldysh space:
\be
\hat G = \begin{pmatrix} G^\text{R} & G^\text{K}\\ 0&G^\text{A} \end{pmatrix},
\qquad
G^\text{K} = G^\text{R}\circ \FF-\FF\circ G^\text{A},
\ee
where $\FF$ is the fermion distribution function, and the symbol ``$\circ$''
denotes the convolution over intermediate spatial and time indices.
At the equilibrium with the temperature $T$, $\FF_\eps=\tanh(\eps/2T)$.

\subsection{Interaction propagator}

In the derivation below we will mainly assume that electrons in the dot interact via the Coulomb potential $V_0(\br)=e^2/\epsilon r$, where $\epsilon$ is the dielectric constant of the medium.
Due to the long-range nature of the Coulomb interaction, $V_0^{-1}(\bq\to 0)=0$,
the propagator $\hat V$ should be calculated taking screening into account
(this procedure is not required for an arbitrary weak and not long-range potential).
As usual, we do this in the random phase approximation (RPA) \cite{Mahan} justified in the case of a small gas parameter:
\be
\label{rs}
  r_s = \frac{e^2}{\epsilon v_F} \ll 1 .
\ee
Under this condition the ratio of the Hartree and Fock diagrams, $\frs$, introduced in Eq.~(\ref{f-gen}) is small too, see Eq.~(\ref{f-Coulomb}). Then one should sum only the bubble contributions to the effective propagator, with independent averaging over disorder in each bubble.

The RPA-screened
interaction propagator has the standard structure:
\be
  \hat V = \begin{pmatrix} V^\text{K} & V^\text{R} \\ V^\text{A} & 0 \end{pmatrix},
\qquad
  V^\text{K}=V^\text{R}\circ \BB-\BB\circ V^\text{A},
\label{CProp}
\ee
where $\BB$ is the boson distribution function defined as \cite{KA99}
\be
\BB_{\omega}=\frac 1{2\omega}\int_{-\infty}^{\infty}(1-\FF_{\ve'}\FF_{\ve'-\omega})d\ve' .
\ee
At the thermal equilibrium, $\BB_\omega=\coth\left(\omega/2T\right)$.

\begin{figure}
\centerline{\includegraphics[width=120mm]{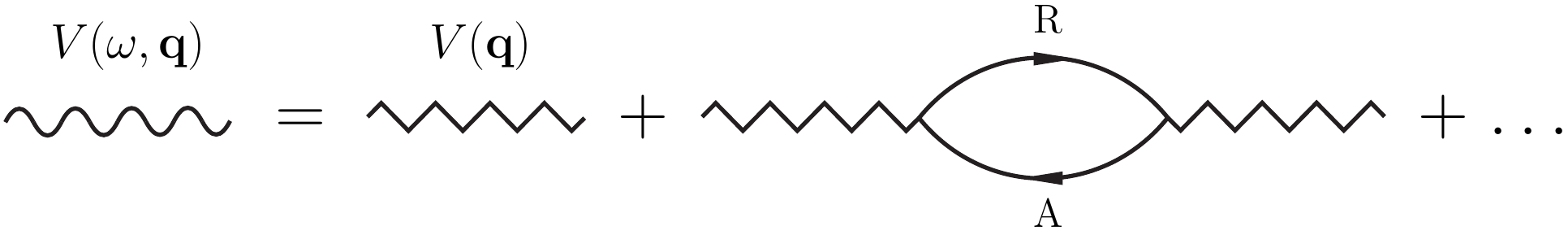}}
\caption{
The dynamically screened interaction $V(\omega,\bq)$ (wavy line) expressed in terms
of the statically screened interaction (SSI) $V(\bq)$ (zigzag line)
and the dynamic polarization operator $G^\text{R}G^\text{A}$.
For a quantum-dot at $(\eps,T)\ll\ETh$, each dynamic polarization bubble contains a small factor of $\omega/\ETh$.}
\label{F:SSI}
\end{figure}

The dynamically screened interaction can be written in the form (see Fig.~\ref{F:SSI})
\be
  V^{\text{R}(\text{A})}(\omega,\bq)
  =
  \bigl[
    V^{-1}(\bq) + \Pi^{\text{R}(\text{A})}_\text{dyn}(\omega,\bq)
  \bigr]^{-1} ,
\label{CProp10}
\ee
where the real function $V(\bq)$ is the statically screened interaction (SSI):
\be
  V(\bq)
  =
  \bigl[
    V_0^{-1}(\bq) + \ns\nu
  \bigr]^{-1} ,
\label{V-SSI}
\ee
and $\Pi_\text{dyn}$ is the dynamic part of the polarization operator originating from the bubble $G^\text{R}G^\text{A}$:
\be
  \Pi^{\text{R}(\text{A})}_\text{dyn}(\omega,\bq)
  =
  \ns\nu \frac{\pm i\omega}{Dq^2\mp i\omega}
  .
\label{Pi10}
\ee
The simple form of the polarization operator $\Pi(\bq,\omega)$ in Eqs.~(\ref{V-SSI}) and (\ref{Pi10}) corresponds to the limit $q\ll p_F$ (and $\omega\ll E_F$). In the case of the Coulomb interaction with a small $r_s$, the spatial dispersion of $\Pi(\bq,0)$ is irrelevant as the interaction in Eq.~(\ref{f-gen}) is determined by $q\sim\kappa\ll p_F$. In the case of a weak short-scale interaction, $\lambda\ll1$, static screening can be neglected and the form of $\Pi(\bq,0)$ is not important.

The SSI defined in Eq.~(\ref{V-SSI}) can be conveniently rewritten in the form [Fourier transform of Eq.~(\ref{V(r)})]
\be
\label{V-static}
  V(\bq)
  = \frac{\lambda}{\ns\nu} v_\kappa(\bq)
 ,
\ee
where
$v_\kappa(\bq)$
is the momentum
representation of the $\delta$ function smeared over the interaction radius, $\delta_\kappa(\br)$.
By definition, $v_\kappa(0)=1$.
In the important case of the Coulomb interaction, $\lambda^\text{Coul}=1$ and
\be
\label{delta-kappa-Coul}
  v_\kappa^\text{Coul}(\bq)
  =
  \begin{cases}
    \kappa_\text{3D}^2/(q^2+\kappa_\text{3D}^2), & \text{in 3D,} \\
    \kappa_\text{2D}/(q+\kappa_\text{2D}), & \text{in 2D,}
  \end{cases}
\ee
where $1/\kappa$ is the Thomas-Fermi screening length given by
\be
\label{kappa}
  \kappa^2_\text{3D} = 4\pi \ns\nu_3 e^2/\epsilon,
\qquad
  \kappa_\text{2D} = 2\pi \ns\nu_2 e^2/\epsilon .
\ee

\subsection{Kinetic equation}

In this Section we sketch the main steps in the derivation of the quantum kinetic equation
in the Keldysh formalism (for a detailed discussion see, e.g, Refs.~\cite{RS,Altshuler}).
From the Dyson equation for the exact (disorder-dependent)
electron Green function $\hat G$ we obtain
\be
  \bigl[\hat G_0^{-1},\hat G\bigr]=\bigl[\hat{\Sigma},\hat G\bigr],
\label{comm}
\ee
where $[A,B]=A\circ B-B\circ A$,
and $\hat{\Sigma}$ is the irreducible self-energy due to interaction,
bearing the same triangular structure as the electron Green function:
\be
\label{Sigma-RAK}
  \hat{\Sigma}=\begin{pmatrix} \Sigma^\text{R}& \Sigma^\text{K} \\ 0& \Sigma^\text{A} \end{pmatrix}.
\ee

As we are interested in the slow dynamics of the system,
it is convenient to switch to the mixed energy-time (Wigner) representation,
\be
  f_{\ve}(t) = \int d\tau \, f\left(t+\frac {\tau}2, t-\frac{\tau}2\right)e^{i\ve \tau} ,
\ee
which allows us to get rid of the convolution on the right-hand side
of Eq.~(\ref{comm}).
The kinetic equation is obtained from the Keldysh component of Eq.~(\ref{comm})
by tracing over the space.
We also assume that the distribution function does not depend
on the coordinates. Therefore the left-hand side of Eq.~(\ref{comm})
produces only the time derivatives of $\FF$, and we arrive at
the standard form of the kinetic equation,
\be
\label{F-St}
  \partial_t \FF_\eps = \St[\FF_\eps] .
\ee
The collision integral $\St[\FF_\eps]$ is given by
\be
  \St[\FF_\eps]
  \int d\br \, \Delta G_{\ve}(\br,\br)
  =
  -i\int d\br \, d\br'\left\{
  \Delta \Sigma_{\ve}(\br, \br')G^\text{K}_{\ve}(\br',\br)
- \Sigma^\text{K}_{\ve}(\br,\br')\Delta G_{\ve}(\br',\br)
  \right\},
\label{kinur0}
\ee
where all functions have an implicit central-time argument $t$, and we denote
\be
  \Delta G= G^\text{R}-G^\text{A},\qquad \Delta \Sigma=\Sigma^\text{R}-\Sigma^\text{A}.
\label{DeltaG}
\ee

\subsection{Modification of the Keldysh technique}
\label{SS:Modified}

As our future analysis will involve diagrams with many interaction propagators,
it is convenient to modify the standard Keldysh diagrammatic technique
discussed above to make it suitable for routine calculations.
We find it appropriate to `eliminate' the Keldysh component
of the electron Green function and to describe the system
in terms of the retarded and advanced Green functions only.
To this end, we note that in the case of weak nonequilibrium
one has $G^\text{K}_{\ve}(t)=\FF_{\ve}(t)\Delta G_{\ve}(t)$,
which allows us to diagonalize the Green function as
\be
\hat G=U^{-1}_{\ve} g_{\ve} U_{\ve},
\qquad
g_{\ve}=\begin{pmatrix} G^\text{R}_{\ve}&0\\0&G^\text{A}_{\ve} \end{pmatrix},
\qquad
U_{\ve}=\begin{pmatrix} 1&\FF_{\ve}\\0&-1\end{pmatrix} ,
\label{gtransform}
\ee
where the time argument is suppressed for brevity.
In order to construct the perturbation theory in terms of
the Green function $g$,
it is convenient to include $U_{\ve}$ into the definition
of the interaction vertex:
\be
\Gamma^k(\ve_1,\ve_2)=U_{\ve_1}\gamma^k U^{-1}_{\ve_2}.
\ee
The resulting $\Gamma^k(\ve_1,\ve_2)$ depend on two
energy indices owing to the energy dependence of the distribution function:
\be
\Gamma^1(\ve_1,\ve_2)=\begin{pmatrix} 1&\FF_{\ve_2}-\FF_{\ve_1}\\0&1\end{pmatrix}, \qquad \Gamma^2(\ve_1,\ve_2)=\begin{pmatrix} \FF_{\ve_1}&-1+\FF_{\ve_1}\FF_{\ve_2}\\-1&-\FF_{\ve_2}\end{pmatrix}.
\ee

This modification of the Keldysh technique that will be used below allows us
to simplify calculations with many interaction lines significantly,
since now the electron Green function $g$ is diagonal in the Keldysh space
and does not contain the distribution function. The interaction line is defined now as
\be
\label{Y-def}
Y_{abcd}(\ve,\ve',\omega,\bq)=
\raisebox{-24pt}{
  \begin{picture}(88,53)(-10,0)
    \put(0,10){\includegraphics[scale=0.15]{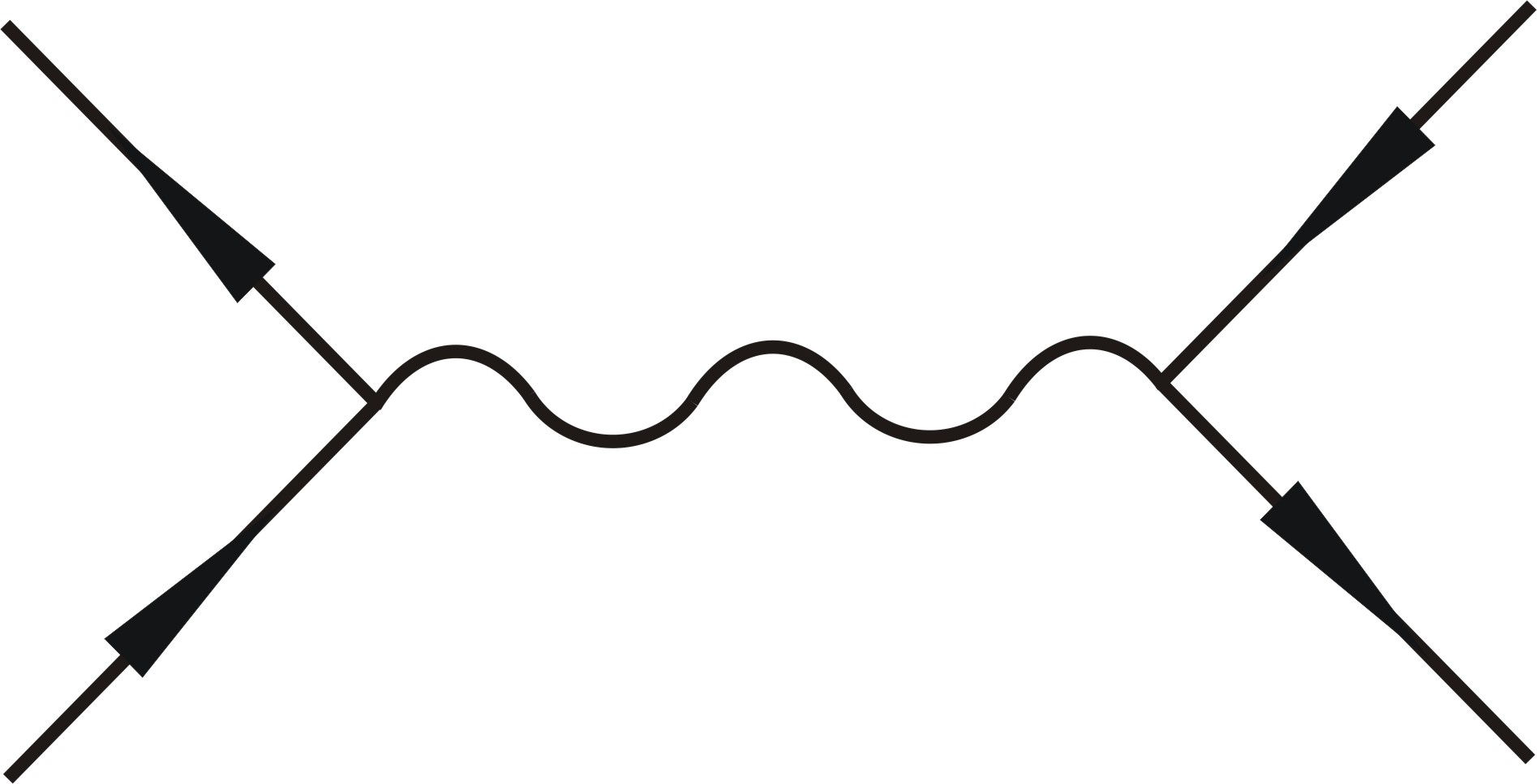}}
    \put(2,2){$\eps$}
    \put(2,48){$\eps-\omega$}
    \put(43,48){$\eps'-\omega$}
    \put(58,2){$\eps'$}
    \put(-5,15){$a$}
    \put(-5,34){$b$}
    \put(68,34){$c$}
    \put(68,15){$d$}
  \end{picture}
  }
  =
  \Gamma^k_{ab}(\ve,\ve-\omega)\Gamma^l_{cd}\left(\ve'-\omega,\ve'\right)V^{kl}_{\omega}(\bq),
\ee
where indices $a,b,c,d$ take the values $\text{R}$ and $\text{A}$,
and we imply the correspondence $\text{R}\leftrightarrow 1$ and $\text{A}\leftrightarrow 2$.
Explicit expressions for $Y_{abcd}$ are listed in \ref{A:Y}.
According to the general rules of the diagrammatic technique,
the element $Y_{abcd}(\ve,\ve',\omega,\bq)$ enters with the coefficient $i/2$
[the factor $2$ is inherited from Eq.~(\ref{propV})].

The irreducible self-energy in the new basis, $\Sigma^{ij}$,
is related to the self-energy (\ref{Sigma-RAK}) as
\be
  \Sigma^\text{R}=\Sigma^{\RR},
\quad
  \Sigma^\text{A}=\Sigma^{\AA},
\quad
  \Sigma^\text{K}=\left(\Sigma^{\RR}-\Sigma^{\AA}\right) \FF_{\ve}-\Sigma^{\RA},
\quad
  \Sigma^{\AR}=0.
\ee
Thus, in the modified Keldysh technique, Eq.~(\ref{kinur0})
for the collision integral takes a compact form:
\be
  \St[\FF_\eps] \int d\br\,\Delta G_{\ve}(\br,\br)
  =
  -i\int d\br\, d\br'\,\Sigma^{\RA}_{\ve}(\br,\br') \Delta G_{\ve}(\br',\br).
\label{kinur1}
\ee

\subsection{Collision integral and the energy relaxation rate}

The inelastic energy relaxation rate, $\gamma(\ve,T)$, can be obtained
from the collision integral in the usual way~\cite{Altshuler,Schmid-ep,Reizer}:
\be
  \gamma(\ve,T) = -\frac{\delta \, \St[\FF_{\ve}]}{\delta \FF_{\ve}}.
\label{tau}
\ee
Using $\St[\FF_{\ve}]$ from Eq.~(\ref{kinur1}), we arrive at the following
expression for $\gamma(\ve,T)$:
\be
  \gamma(\ve,T) \int d\br\,\Delta G_{\ve}(\br,\br)
  =
  i\frac{\delta}{\delta \FF_{\ve}}
  \int d\br \, d\br'\,\Sigma^{\RA}_{\ve}(\br,\br') \Delta G_{\ve}(\br',\br).
\label{gamma}
\ee

Equation (\ref{gamma}) is the starting point for the calculation of the energy relaxation rate.
This equation contains exact disorder-dependent Green functions, and averaging its $n$'th power
over the random potential generates the $n$'th moment of $\gamma(\ve,T)$.
The simplest is the first moment, $\corr{\gamma(\ve,T)}$, related to the average
collision integral $\corr{\St[\FF_{\ve}]}$. In this case, the left-hand side of Eq.~(\ref{kinur1})
gives the average density of states, $\corr{\Delta G_{\ve}(\br,\br)} = -2\pi i \nu$,
and one arrives at
\be
  \corr{\St[\FF_{\ve}]}
  =
  \frac{\Delta}{2\pi} \int d\br\, d\br'
  \langle\Sigma^{\RA}_{\ve}(\br,\br') \Delta G_{\ve}(\br',\br)\rangle .
\label{kinur01}
\ee
In the next Section we show how this equation reproduces Sivan, Imry and Aronov result (\ref{SIAT}) for the average inelastic rate in the limit $F\to0$, and generalize it to the case of an arbitatry parameter $F$.
The second moment of $\gamma(\eps,T)$ will be considered in Secs.~\ref{S:MF}--\ref{S:mesofluct-rs}.

\section{Average energy relaxation rate}
\label{S:Average}

In this Section we rederive the result of Sivan, Imry and Aronov \cite{SIA}
and generalize it to the case
of an arbitrary temperature $T$, spin degeneracy $\ns$ and interaction radius characterized by the parameter $\frs$.
Our treatment closely follows the standard derivation
of the kinetic equation in the RPA approximation \cite{RS}, but within the modified
Keldysh technique introduced in Sec.~\ref{SS:Modified}.
The purpose of this Section is to illustrate the usage of this technique
and to prepare the ingredients for the analysis of mesoscopic fluctuations
of $\gamma(\eps,T)$ in Sec.~\ref{S:MF}.

\subsection{Derivation of the Sivan, Imry and Aronov result}
\label{SS:SIA}

\begin{figure}
\centerline{\includegraphics[width=163mm]{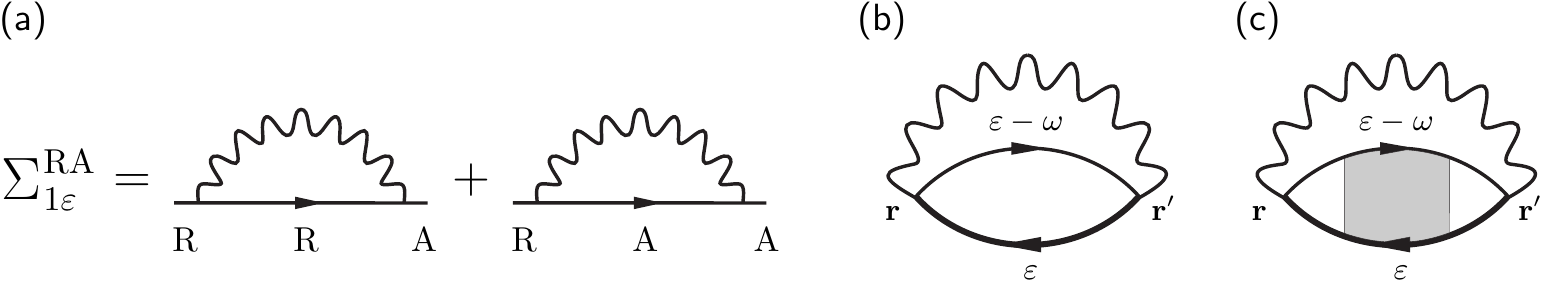}}
\caption{(a) The simplest
(and the most important for small $\frs$) contribution to the self-energy $\Sigma^{\RA}_{\eps}$.
(b) The graphic representation of the product $\Sigma^{\RA}_{\ve} \Delta G_{\ve}$
which determines the collision integral (\ref{kinur01}). The thick line
standing for $\Delta G_{\ve} = G^\text{R}_{\eps}-G^\text{A}_{\eps}$ corresponds to
the electron state whose decay is studied.
(c) The diagram (b) averaged over disorder, with the shaded block representing the diffuson.}
\label{F:SIA2}
\end{figure}

The simplest diagram for the self-energy $\Sigma^{\RA}_{\ve}$
is shown in Fig.~\ref{F:SIA2}(a).
In the limit $\frs\ll1$ it gives the leading contribution
to the relaxation rate (a more general situation is discussed
in Sec.~\ref{SS:rs}).
The corresponding analytic expression
can be easily written in terms of the elements $Y$ introduced in Eq.~(\ref{Y-def}):
\be
\Sigma^{\RA}_{1\ve}(\br,\br')=\frac i2\int (d\omega)\left\{G^\text{R}_{\ve-\omega}(\br,\br')Y_{\RR,\RA}(\ve,\ve,\omega, \br-\br')+G^\text{A}_{\ve-\omega}(\br,\br')Y_{\RA,\AA}(\ve,\ve,\omega,\br-\br')\right\},
\ee
with $(d\omega)\equiv d\omega/2\pi$.
The interaction propagators are assumed to be RPA screened
and averaged over disorder (see Sec.~\ref{SS:Keldysh}),
while the electron Green functions are still exact in a given realization of disorder.
Using Eqs.~(\ref{RRRA}) and (\ref{RAAA}), and employing the analyticity properties
we obtain
\be
\Sigma^{\RA}_{1\ve}(\br,\br')
=
\frac i2 \int (d\omega) \Phi(\ve,\omega)
\Delta G_{\ve-\omega}(\br,\br')
\Delta V(\omega, \br-\br'),
\label{Sigma1}
\ee
where $\Delta V=V^\text{R}-V^\text{A}$, and
\be
  \Phi(\ve,\omega)=(\FF_{\ve}-\FF_{\ve-\omega})\BB_{\omega}-(1-\FF_{\ve}\FF_{\ve-\omega})
\ee
is a combination of the fermionic and bosonic distribution functions which vanishes at the equilibrium (detailed balance).

The collision integral (\ref{kinur01}) involves the product $\Sigma^{\RA}_{\ve} \Delta G_{\ve}$,
which is to be averaged over disorder. To make this calculation more intuitive,
we represent it diagrammatically on a separate graph in Fig.~\ref{F:SIA2}(b),
where the thick line stands for the Green function $\Delta G_{\ve}$
of the initial state. To avoid confusion we emphasize that this picture is just
a simple graphical notation for $\Sigma^{\RA}_{\ve} \Delta G_{\ve}$, since
the vertices with a thick line are not described by the rule of Eq.~(\ref{Y-def}).
Substituting the self-energy (\ref{Sigma1}) into Eq.~(\ref{kinur01})
we obtain for the averaged collision integral:
\be
  \corr{\St[\FF_{\ve}]}
  =
  \frac{i\Delta}{4\pi}\int d\br \,d\br'(d\omega) \Phi(\ve,\omega)
  \langle\Delta G_{\ve}(\br',\br) \Delta G_{\ve-\omega}(\br,\br')\rangle
  \Delta V(\omega,\br-\br') .
\ee
The averaged product of two Green functions,
\be
\label{<GG>}
  \corr{\Delta G_{\ve}(\br',\br) \Delta G_{\ve-\omega}(\br,\br')}
  =
  -4\pi\nu \Re D_0^\text{R}(\omega,\br-\br')
\ee
is expressed in terms
of the particle-hole ladder (diffuson), $D_0^\text{R}(\omega,\bq) = 1/(Dq^2-i\omega)$,
see Fig.~\ref{F:SIA2}(c). Then one arrives at the well-known result for the collision integral in dirty metals (see, for example, Refs.~\cite{RS,AltshulerAronov,Schmid-ee}):
\be
  \corr{\St[\FF_\ve]}
  =
  2\int(d\bq)(d\omega)\Phi(\ve,\omega)\Re D_0^\text{R}(\omega,\bq)\Im V^\text{R}(\omega,\bq) ,
\label{StF}
\ee
where $(d\bq)\equiv d^dq/(2\pi)^d$, and $d$ is the space dimensionality.

According to Eq.~(\ref{CProp10}), the imaginary part of the fluctuation propagator is determined by the dynamic polarization operator $\Pi_\text{dyn}$.
For a quantum dot in the zero-dimensional regime ($\omega\ll\ETh$),
$\Pi_\text{dyn}$ is a small correction to $V^{-1}(\bq)$,
and $\Im V^\text{R}(\omega,\bq)$ can be written as
\be
\label{ImV}
  \Im V^\text{R}(\omega,\bq)
  \approx
  - V^2(\bq) \Im \Pi^{\text{R}}_\text{dyn}(\omega,\bq) .
\ee
Using Eq.~(\ref{Pi10}) and assuming that the system size, $L$, is larger than the interaction radius, $1/\kappa$, we obtain in the zero-dimensional regime:
\be
  \corr{\St[\FF_{\ve}]}
  =
  -\frac{2\lambda^2}{\ns\nu}\int (d\omega) (d\bq)\frac{\omega \, \Phi(\ve,\omega)}{(Dq^2)^2}
  =
  - \frac{2\lambda^2\Delta}{\ns} \mathop{{\sum}'}_m
  \int (d\omega)\frac {\omega\,\Phi(\ve,\omega)}{(Dq_m^2)^2} ,
\label{StF1}
\ee
where $q_m^2$ are the eigenvalues of the operator $-\nabla^2$ in the dot
with von Neumann boundary conditions, with the zero mode being excluded
due to electroneutrality \cite{SIA,Blanter,AGKL,AG98,BlanterMirlin1997}.
Performing summation over discrete momenta, we get
\be
  \corr{\St[\FF_{\ve}]}
  =
  - \frac{2\lambda^2\Delta}{\ns E_2^2}
  \int (d\omega) \, \omega \, \Phi(\ve,\omega) ,
\ee
where the energy $E_2\sim\ETh$ is defined in Eq.~(\ref{En}).
The inelastic energy relaxation rate at the equilibrium
can be extracted with the help of Eq.~(\ref{tau}):
\be
\label{gamma0-integral}
\gammaRPA(\ve,T)=\frac{2\lambda^2\Delta}{\ns E_2^2} \int (d\omega)\,\omega \left\{ \coth\left(\frac{\omega}{2T}\right)+\tanh\left(\frac{\ve-\omega}{2T}\right)\right\}.
\ee
Integrating over $\omega$, we arrive at Eq.~(\ref{SIAT})
for the average energy relaxation rate $\gamma_0(\ve,T)$
in the limit $\frs\to0$.

\begin{figure}
\centerline{\includegraphics[width=150mm]{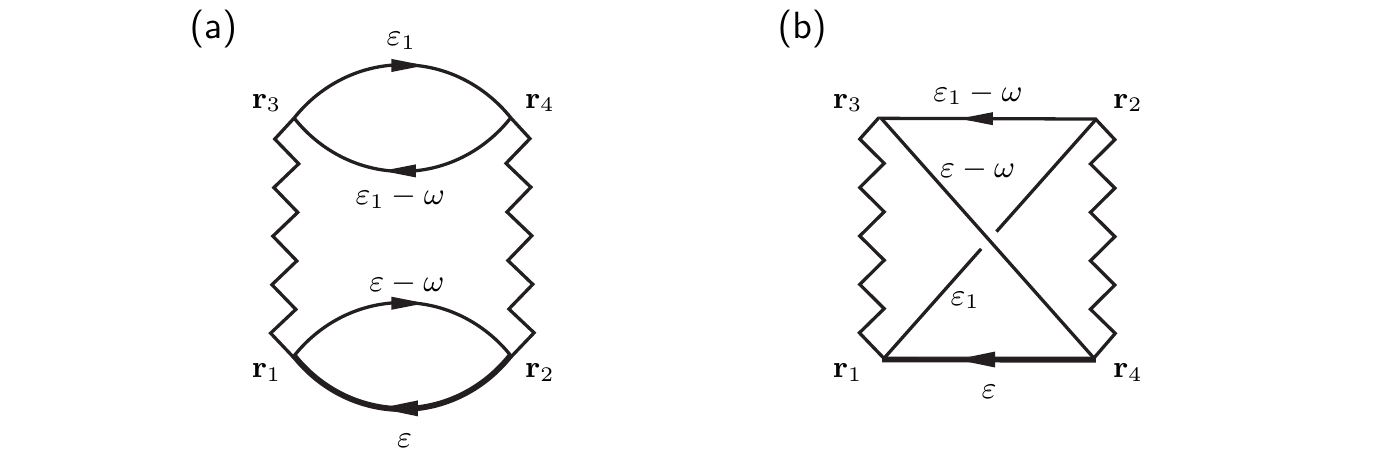}}
\vspace{0mm}
\caption{
The diagrams for the collision integral with two SSI lines
prior to disorder averaging.
(a) The RPA diagram
\ref{F:SIA2}(b)
redrawn in terms of the SSI lines (\ref{V-static})
and the dynamic polarization bubble $G^\text{R}G^\text{A}$. The corresponding contribution to the relaxation rate is given by the $X_4$ term in Eq.~(\ref{gamma1}).
(b) The other, non-RPA diagram discarded in Sec.~\ref{SS:SIA}, leading to the $Y_4$ term in Eq.~(\ref{gamma1}). Its contribution to the mean relaxation rate can be neglected
at small $\frs$ [see Eq.~(\ref{gamma1full})].}
\label{F:2Coulomb}
\end{figure}

\subsection{Physical interpretation and the other diagram}
\label{SS:Physical}

It is instructive to rederive the result for $\gammaRPA(\ve,T)$
in a slightly different manner to clarify the physics of the decay
process responsible for the width of an one-electron level.
Real decay processes are described by $\Im V^\text{R}(\omega,\bq)$ [see Eq.~(\ref{StF})],
which is determined by the dynamic screening of the bare interaction.
In the zero-dimensional limit, $(E,T,\omega)\ll\ETh$,
the dynamic part of the polarization bubble,
$\Pi_\text{dyn}(\omega,\bq)$, contains
a small factor $\omega/\ETh$ [see Eq.~(\ref{Pi10})].
This means that with the accuracy of order $\omega/\ETh$ we may
consider $\Pi_\text{dyn}(\omega,\bq)$ as a small perturbation
and retain only
the first dynamic bubble in the expansion of $V(\omega,\bq)$
near the statically screened interaction (SSI)
$V(\bq)$, see Eq.~(\ref{ImV}).

Therefore with the accuracy of order $\omega/\ETh$ the diagram in Fig.~\ref{F:SIA2}(a)
for the average collision integral with one dynamically screened interaction
can be redrawn as the diagram in Fig.~\ref{F:2Coulomb}(a)
with two SSI lines.
The advantage of the diagrammatic representation of Fig.~\ref{F:2Coulomb}(a)
is that it elucidates the physics of the inelastic collision:
The cross-section of this diagram corresponds to the decay of an electron with
energy $\ve$ into an electron with energy $\ve-\omega$ and an electron-hole
pair with energies $\ve_1$ and $\ve_1-\omega$.
The corresponding contribution to the self-energy
(before disorder averaging)
can be easily read off from Eq.~(\ref{Y-def}):
\be
\label{SigmaRAa}
\Sigma^{\RA}_{a}=-\ns\left(\frac i2\right)^2\sum_{abc}G^a_{\ve-\omega}(\br_1,\br_2)G^b_{\ve_1-\omega}(\br_4,\br_3)G^c_{\ve_1}(\br_3,\br_4) Y_{\text{R}abc}(\ve,\ve_1,\omega,\br_1,\br_3)Y_{cba\text{A}}(\ve_1,\ve,\omega,\br_2,\br_4),
\ee
where here and in what follows the interaction line $Y_{abcd}$ corresponds to the SSI $V(\bq)$.
The factor $-\ns$ in Eq.~(\ref{SigmaRAa}) comes from the upper closed electron loop [the bottom bubble containing a highlighted Green function at the external energy $\Delta G_{\eps}$
is not a closed loop in the diagrammatic sense (see Fig.~\ref{F:SIA2}) and does not contribute an extra factor $-\ns$].

The diagram \ref{F:2Coulomb}(a) is not the only one with two SSI lines
that describes quasiparticle decay process. The other diagram is shown in Fig.~\ref{F:2Coulomb}(b). Though it is not as intuitive as the diagram \ref{F:2Coulomb}(a), it also makes a contribution to the relaxation rate. That contribution is usually discarded
since it vanishes in the limit $\frs\to0$ (see below).
However as we show in Sec.~\ref{Seconddiagram},
the diagram \ref{F:2Coulomb}(b) should be taken into account
in calculating mesoscopic fluctuations of the relaxation rate,
since its contribution is comparable to that of the diagram \ref{F:2Coulomb}(a)
even in the limit $\frs\to0$.
The self-energy for the diagram \ref{F:2Coulomb}(b) is given by
\be
\Sigma^{\RA}_{b}=\left(\frac i2\right)^2\sum_{abc}G^a_{\ve_1}(\br_1,\br_2)G^b_{\ve_1-\omega}(\br_2,\br_3)G^c_{\ve-\omega}(\br_3,\br_4) Y_{\text{R}abc}(\ve,\ve-\omega,\ve-\ve_1,\br_1,\br_3)Y_{abc\text{A}}(\ve_1,\ve,\omega,\br_2,\br_4).
\ee
Due to the absence of the closed electron loop in the diagram \ref{F:2Coulomb}(b), $\Sigma^{\RA}_{b}$ does not contain an extra factor $-\ns$ compared to $\Sigma^{\RA}_{a}$.

The energy relaxation rate $\gamma(\ve, T)$ (in a given realization of disorder) due to processes with two SSI
lines can be obtained from Eq.~(\ref{gamma}) with $\Sigma^{\RA}=\Sigma^{\RA}_{a}+\Sigma^{\RA}_{b}$.
After some algebra involving analyticity properties we obtain
\be
  \gamma(\ve,T)\int d\br \, \Delta G_{\ve}(\br,\br)
  =
  - \frac{i\lambda^2}{4\ns^2\nu^2}\int d\br_1\ldots d\br_4\int (d\ve_1)(d\omega)(\FF_{\ve_1}-\FF_{\ve_1-\omega})(\FF_{\ve-\omega}+\BB_{\omega})
  \left[ \ns X_4-Y_4 \right],
\label{gamma1}
\ee
where
\be
\label{X4}
  X_4
  =
  \delta_\kappa(\br_1-\br_3)\delta_\kappa(\br_2-\br_4) \,
  \Delta G_{\ve}(\br_2,\br_1)\Delta G_{\ve-\omega}(\br_1,\br_2)
  \Delta G_{\ve_1-\omega}(\br_4,\br_3)\Delta G_{\ve_1}(\br_3,\br_4)
\ee
and
\be
\label{Y4}
Y_4=\delta_{\kappa}(\br_1-\br_3)\delta_{\kappa}(\br_2-\br_4)\Delta G_{\ve}(\br_4,\br_1)\Delta G_{\ve_1}(\br_1,\br_2)\Delta G_{\ve_1-\omega}(\br_2,\br_3)\Delta G_{\ve-\omega}(\br_3,\br_4).
\ee
In Eq.~(\ref{gamma1}), the term $\ns X_4$ comes from $\Sigma^{\RA}_{a}$ [diagram \ref{F:2Coulomb}(a)], and the term $-Y_4$ comes from $\Sigma^{\RA}_{b}$ [diagram \ref{F:2Coulomb}(b)].
An expression similar to Eq.~(\ref{gamma1}) has been derived in Ref.~\cite{Basko}, where only the contribution from the
diagram \ref{F:2Coulomb}(a) has been considered.
In the limit when the interaction radius $1/\kappa$ exceeds the Fermi wavelength, i.e. at $\frs\ll1$, disorder averaged $\langle \gamma(\ve,T)\rangle$
reproduces the result (\ref{gamma0-integral}) for $\gammaRPA(\ve, T)$ [see Eq.~(\ref{gamma1full})].

\begin{figure}
\centerline{\includegraphics[width=150mm]{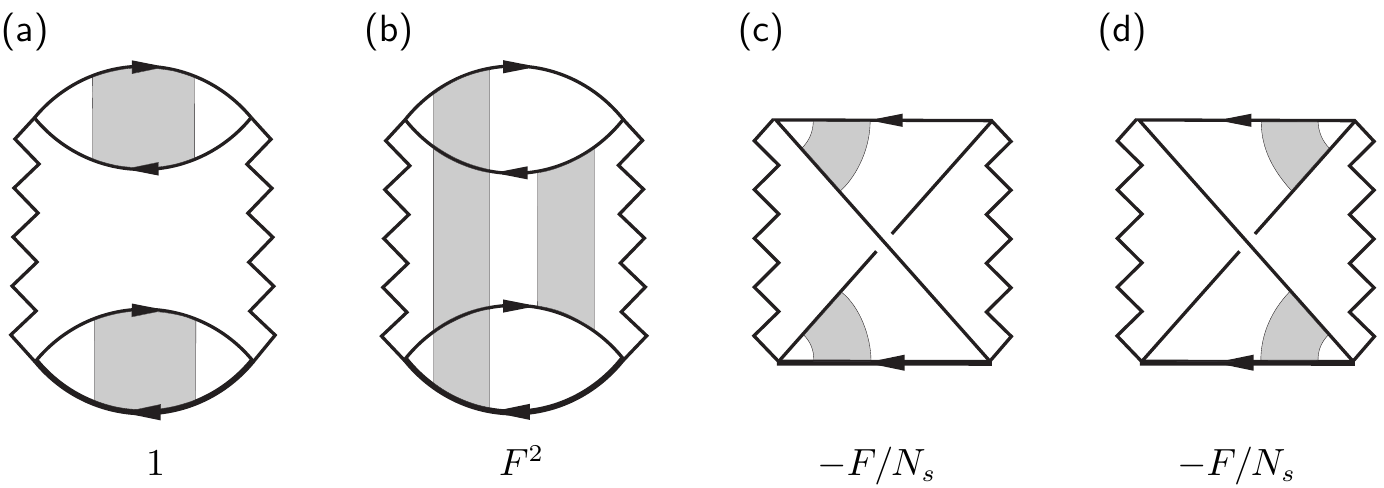}}
\caption{Possible ways to average the diagrams for the relaxation rate shown in Fig.~\ref{F:2Coulomb} over disorder (other types of averaging with two diffusons exist but are suppressed at least by the factor $l/L\ll1$, where $l$ is the mean free path).
(a) RPA averaging of the diagram \ref{F:2Coulomb}(a)
reproducing the SIA result for $\gammaRPA(\varepsilon,T)$ [Eq.~(\ref{gamma0-integral})].
(b) Non-RPA averaging of the diagram \ref{F:2Coulomb}(a)
with the contribution $\frs^2\gammaRPA(\varepsilon,T)$.
(c) and (d) Two ways to average the diagram \ref{F:2Coulomb}(b), each producing the contribution $-(\frs/\ns)\gammaRPA(\varepsilon,T)$ to the relaxation rate.}
\label{F:2Coulomb-av}
\end{figure}

\subsection{Disorder averaging and the role of the parameter $\frs$}
\label{SS:rs}

Equation (\ref{gamma1}) is the starting point for evaluating moments of the relaxation rate. In order to find its mean
value one has to average the combinations of four Green functions in Eqs.~(\ref{X4}) and (\ref{Y4}).

The RPA result of Sec.~\ref{SS:SIA} is reproduced if one takes only the diagram \ref{F:2Coulomb}(a) (the term $X_4$) into account and average each bubble independently [see Fig.~\ref{F:2Coulomb-av}(a)]:
\be
\label{X4RPA}
  \corr{X_4}_\text{RPA}
  =
  \delta_\kappa(\br_1-\br_3)\delta_\kappa(\br_2-\br_4)
  \corr{\Delta G_{\ve}(\br_2,\br_1)\Delta G_{\ve-\omega}(\br_1,\br_2)}
  \corr{\Delta G_{\ve_1-\omega}(\br_4,\br_3)\Delta G_{\ve_1}(\br_3,\br_4)} .
\ee
Averaging with the help of Eq.~(\ref{<GG>}) and integrating
over $\eps_1$, one arrives at Eq.~(\ref{gamma0-integral}) for $\gammaRPA(\ve,T)$.

Along with the RPA averaging (\ref{X4RPA}), there exists another way to average $X_4$ over disorder shown in Fig.~\ref{F:2Coulomb-av}(b). It has the same structure of diffusons as the RPA diagram \ref{F:2Coulomb-av}(a),
but with the interaction lines taken at fast momenta $q\sim\min(\kappa,p_F)$.
The corresponding contribution to the relaxation rate, $\frs^2\gammaRPA(\ve,T)$,
differs from the RPA-contribution by a factor $\frs^2$, where $\frs$ is the standard notation for the ratio of the Hartree and Fock diagrams \cite{Altshulerbook,Akkermans,AAL} discussed in \ref{S:Non-RPA}.

Finally, consider disorder averaging of the diagram \ref{F:2Coulomb}(b) corresponding to the term $Y_4$ in Eq.~(\ref{gamma1}). For this diagram, there are also two ways to draw two-diffuson configurations shown in Figs.~\ref{F:2Coulomb-av}(c) and (d).
Their contributions are equal and can be calculated similar to that of the
diagram \ref{F:2Coulomb-av}(b), but now only one of the two SSI lines are taken at fast momentum making the result proportional to the first power of $\frs$.
The overall correction of the diagram \ref{F:2Coulomb}(b) to the average relaxation rate
is given by $-(2\frs/\ns)\gammaRPA(\ve, T)$.

Thus we see that among four possible contributions to the
average relaxation rate shown in Fig.~\ref{F:2Coulomb-av},
all non-RPA diagrams [(b), (c) and (d)] are proportional to some
power of $\frs$ and hence are suppressed
if the interaction radius is larger than the Fermi wavelength.
The smallness of the non-RPA diagrams
in this limit should be attributed to the Friedel oscillations
which suppress the contribution of the corresponding process \cite{Mahan}.
Such a situation was considered, e.g., in Refs.~\cite{RS,AltshulerAronov,Schmid-ee}.
On the other hand, for the point-like interaction (corresponding to $\frs=1$),
the diagrams \ref{F:2Coulomb-av}(a) and \ref{F:2Coulomb-av}(b) give the same
contribution. This short-range limit was considered in
Refs.~\cite{Blanter,AGKL,AG98,ABG}.

Collecting the contributions of all diagrams with two SSI lines
and two diffusons (Fig.~\ref{F:2Coulomb-av}), we obtain the final result for the average energy relaxation rate $\gamma_0(\ve,T)=\corr{\gamma(\ve,T)}$:
\be
\label{gamma1full}
  \gamma_0(\ve, T)
  =
  \left( 1 - 2\frs/\ns + \frs^2 \right) \gammaRPA(\ve,T) ,
\ee
leading to Eq.~(\ref{SIAT}).
Note that for spinless electrons with point-like interaction,
the inelastic relaxation rate vanishes, $\gamma=0$,
which is a consequence of the Pauli exclusion principle.
It is essential that such a cancellation takes place only
if the diagram \ref{F:2Coulomb}(b) is taken into account.

\section{Mesoscopic fluctuations of the energy relaxation rate: general consideration}
\label{S:MF}

In this Section we discuss the general approach to the calculation of mesoscopic fluctuations of the energy relaxation rate $\gamma(\ve,T)$.
The starting point is the diagrams for the collision integral shown
in Figs.~\ref{F:2Coulomb}(a) and \ref{F:2Coulomb}(b).
The corresponding expression for $\gamma(\ve,T)$ is given by
Eq.~(\ref{gamma1}) which is written for a particular realization of impurities. Mesoscopic fluctuations of the relaxation rate are determined by the square of $\gamma(\ve,T)$ averaged over disorder:
\begin{multline}
  \langle \gamma^2(\ve,T)\rangle
  \int d\br\, d\br' \,
  \langle \Delta G_{\ve}(\br,\br)\Delta G_{\ve}(\br',\br')\rangle
  =
- \frac{\lambda^4}{16\ns^4\nu^4}\int d\br_1\ldots d\br_8
  \int (d\ve_1)(d\ve_2)(d\omega_1)(d\omega_2)
\\{}
  \times
  (\FF_{\ve_1}-\FF_{\ve_1-\omega_1})(\FF_{\ve-\omega_1}+\BB_{\omega_1})
  (\FF_{\ve_2}-\FF_{\ve_2-\omega_2})(\FF_{\ve-\omega_2}+\BB_{\omega_2})
  \corr{(\ns X_4-Y_4)(\ns X_4'-Y_4')} ,
\label{corr}
\end{multline}
where the objects $X_4$ and $Y_4$ containing different products of four Green functions and two SSI lines are defined in Eqs.~(\ref{X4}) and (\ref{Y4}), while
$X_4'$ and $Y_4'$ are obtained from them through the replacement
\be
  \{\br_i\} \to \{\br_{i+4}\} ,
\quad
  \ve_1 \to \ve_2 ,
\quad
  \omega_1 \to \omega_2 .
\ee
Equation (\ref{corr}) contains three different products of eight Green functions, $X_4X_4'$, $X_4Y_4'$ and $Y_4Y_4'$, which need to be separately averaged over disorder (the term $Y_4X_4'$ reduces to $X_4Y_4'$ after an obvious change of variables). For example, the explicit form of $X_4X_4'$ is given by
\begin{multline}
  X_4X_4'
  =
  \delta_\kappa(\br_1-\br_3)\delta_\kappa(\br_2-\br_4)
  \delta_\kappa(\br_5-\br_7)\delta_\kappa(\br_6-\br_8)
  \,
  \Delta G_{\ve}(\br_2,\br_1) \Delta G_{\ve-\omega_1}(\br_1,\br_2)
\\{}
  \times
  \Delta G_{\ve_1-\omega_1}(\br_4,\br_3) \Delta G_{\ve_1}(\br_3,\br_4)
  \Delta G_{\ve}(\br_6,\br_5) \Delta G_{\ve-\omega_2}(\br_5,\br_6)
  \Delta G_{\ve_2-\omega_2}(\br_8,\br_7)\Delta G_{\ve_2}(\br_7,\br_8)
  .
\label{X8}
\end{multline}
Since $\Delta G = G^\text{R}-G^\text{A}$, each of the products $X_4X_4'$, $X_4Y_4'$ and $Y_4Y_4'$ contains $2^8=256$ different elementary contributions in terms of $G^\text{R}$ and $G^\text{A}$.

Our task is to calculate the irreducible part $\ccorr{\gamma^2(\ve,T)}$ defined in Eq.~(\ref{corr-irr-def}).
In a usual situation, mesoscopic fluctuations of a random quantity $x$ are determined by irreducible averaging over disorder, when two copies of $x$ are connected by at least one impurity line. In the present case the situation is different as the left-hand side of the basic Eq.~(\ref{corr}) contains the pairwise correlator $\corr{ \Delta G_{\ve} \Delta G_{\ve} }$, which deviates significantly from its reducible part $\corr{\Delta G_{\ve}}^2$ (see Sec.~\ref{paircorr}).
Therefore, even extracting the square of the average, $\corr{\gamma(\ve,T)}^2$, from Eq.~(\ref{corr}) should be done with care as it involves irreducible disorder averaging of
$\corr{(\ns X_4-Y_4)(\ns X_4'-Y_4')}$
needed for non-perturbative account for correlations between
two Green functions with the energy $\ve$. The details of this calculation are presented in \ref{reducible part}.
Such a complication is the price one has to pay for extracting the property of a single discrete level with the technique well suited for describing continuous spectra.
Keeping that in mind we proceed with evaluation of the irreducible part of $\corr{\gamma^2(\ve,T)}$.

In order to determine $\corr{\gamma^2(\ve,T)}$ one has to compute
$\corr{X_4X_4'}$, $\corr{X_4Y_4'}$ and $\corr{Y_4Y_4'}$,
and evaluate four remaining energy integrals in Eq.~(\ref{corr}).
This calculation is rather nontrivial and will be performed in
Secs.~\ref{S:nonpert}, \ref{S:final} and \ref{S:mesofluct-rs}.
Meanwhile we discuss the main idea and outline the principal technical steps of the derivation.

The most complicated task of this procedure is to perform disorder averaging which becomes rather involved in the low-energy limit, $(\eps,T)\ll\ETh$, we are interested in.
Indeed, the single-particle spectrum is well resolved in this case, $\corr{\gamma(\eps,T)}\ll\Delta$, indicating that the averaging is to be performed non-perturbatively.
That can be achieved with the help of the Efetov's nonlinear supersymmetric sigma model technique \cite{Efetov-book}, properly generalized to the case of several Green functions with different energies.

The averaging of two Green functions on the left-hand side of Eq.~(\ref{corr}) is straightforward and can be done exactly in the zero-dimensional geometry, see Sec.~\ref{paircorr}.

The calculation of
$\corr{X_4X_4'}$, $\corr{X_4Y_4'}$ and $\corr{Y_4Y_4'}$,
containing eight Green functions is much more complicated and cannot be performed exactly in the general case.
Fortunately, the exact expression is not required for determination of the leading contribution to mesoscopic fluctuations of the relaxation rate.
A significant simplification is suggested by analysing the physics of the decay process.
According to the FGR, $\gamma(\ve, T)$ can be considered as a sum of the decay processes of a single-particle excitation into all possible final three-particle states allowed by the energy conservation:
$\ve\to(\ve-\omega,\eps',-\eps'+\omega)$. Consequently, $\gamma^2(\ve, T)$ given by Eq.~(\ref{corr}) contains,
in principle, six different final states: $(\ve-\omega_1,\eps_1,-\eps_1+\omega_1, \ve-\omega_2,\eps_2,-\eps_2+\omega_2)$.

However, as it was first demonstrated in Ref.~\cite{BAA}, the leading contribution to mesoscopic fluctuations of the relaxation rate comes from those configurations that describe \emph{the square of the same decay process}, i.e., from the terms with the identical set of the final states.
Such a situation can be realized with two choices:
\be
\label{energy-matching}
  \text{(a) \: $\ve_1 \approx \ve_2$ and $\omega_1 \approx \omega_2$},
\qquad
  \text{(b) \: $\ve_1 \approx \ve-\omega_2$ and $\ve_2 \approx \ve-\omega_1$} ,
\ee
where the energies should coincide with the accuracy of the single level width $\gamma$.
Technically, the importance of such configurations is related to the appearance of the formally divergent delta function of zero frequency, $\delta(0)$, if conditions (\ref{energy-matching}) are considered as strict equalities, corresponding to the use of non-interacting Green functions in
$\corr{X_4X_4'}$, $\corr{X_4Y_4'}$ and $\corr{Y_4Y_4'}$.
In order to take into account the single-particle level broadening due to interaction one should add an imaginary part to the energy argument $E$ of the Green function, replacing it by $E_+$ (for $G^\text{R}$) and $E_-$ (for $G^\text{A}$):
\be
\label{width}
  E_\pm = E \pm i\gamma(E)/2 ,
\ee
where $\gamma(E) \equiv \gamma_0(E,T)$ is the average value of the relaxation rate obtained in the FGR approximation [Eq.~(\ref{SIAT})].
The substitution (\ref{width}) is equivalent to including the elastic part of the electron-electron interaction in the zero-dimensional diffuson (see Sec.~\ref{slowaveraging} and \ref{eldiffuson}).
A finite imaginary part cuts the singularity in $\delta(0)$ which should be replaced roughly by $1/\gamma$, making fluctuations
finite but divergent as $\ve$ and $T$ are decreased.
This enhancement of fluctuations at low energies and temperatures is precisely the effect we are looking for.

Such a mechanism of enhancement of mesoscopic fluctuations of the inelastic width with decreasing temperature was suggested in Ref.~\cite{BAA} for a model when the matrix elements of the interaction $V_{\alpha\beta\gamma\delta}$ in the basis of exact single-particle states were assumed to be independently distributed Gaussian variables.
Our task is more complicated as we do not make any assumptions about the structure of the matrix elements in Fock space and calculate them for the real Coulomb interaction. Though we do not use the language of the matrix elements $V_{\alpha\beta\gamma\delta}$, they are effectively generated after proper averaging over fast diffusive modes
\cite{Blanter,AGKL}.

Having discussed the main idea, we now outline the crucial steps in the calculation of
$\corr{X_4X_4'}$, $\corr{X_4Y_4'}$ and $\corr{Y_4Y_4'}$:
\begin{itemize}
\item
Each of $X_4X_4'$, $X_4Y_4'$ and $Y_4Y_4'$ contains in general $2^8=256$ elementary products of $G^\text{R}$ and $G^\text{A}$.
As we discuss in Secs.~\ref{slowaveraging} and \ref{qual},
the principal contribution to $\ccorr{\gamma^2(\eps,T)}$ originates from the terms with four $G^\text{R}$ and four $G^\text{A}$. Each of these  $C_8^4=70$ terms
may be presented as a functional integral over the 16-component superfields which is then averaged over disorder following the standard technique \cite{Efetov-book}.
The resulting supersymmetric nonlinear sigma model is formulated in terms of the functional integral over the $16\times16$ superfield $Q(\br)$, see Sec.~\ref{SS:SUSY}.

\item
As the SSI lines carry nonzero momenta (otherwise the diagram is zero due to electroneutrality), momentum conservation requires the presence of some number of fast ($q\neq0$) diffusive modes.
Since each fast diffuson contributes a small factor of $\Delta/\ETh$, their number should be minimized.
For $\corr{\gamma(\ve,T)}$ the minimal number was two [see Fig.~\ref{F:2Coulomb-av}], and for $\corr{\corr{\gamma^2(\ve,T)}}$ it is four, producing the
factors $1/E_2^4$ and $1/E_4^4$.
Following the method developed in Ref.~\cite{KM}, we integrate over these fast modes perturbatively and derive an effective action for the zero-mode (spatially-uniform supermatrix $Q$). This procedure is described in details in Sec.~\ref{fastaveraging}, with the total number of emergent contributions $7!!=105$
for every given elementary term from $X_4X_4'$, $X_4Y_4'$ and $Y_4Y_4'$ with four $G^\text{R}$ and four $G^\text{A}$.

\item
The resulting zero-dimensional sigma model is still too complicated
because of the large size of the $Q$ matrix.
However, as described above, the leading contribution comes from configurations
where among energy arguments of eight Green functions, four are pairwise equal
(with an uncertainty of $\gamma$).
Since the typical energy difference between pairs is $(\ve,T)\gg\mls$,
our extended sigma model splits into four blocks, each corresponding to the standard Efetov sigma model for $\corr{G^\text{R}G^\text{A}}$.
The integrals over these sigma models are then evaluated non-perturbatively using the standard technique \cite{Efetov-book}.

The detailed discussion of this step in connection with the choice of pairs is presented in Secs.~\ref{slowaveraging} and \ref{SS:further}.

\item
As a result of the described procedure, quite a few terms will be generated.
Fortunately, most of them will be either zero or less singular than expected.
Only a small number of terms will effectively contribute to fluctuations of
the relaxation rate.
This selection is discussed in Sec.~\ref{S:final} in the simplest case of $\frs\to0$.
The final evaluation of $\ccorr{\gamma^2(\ve,T)}$ in the general case of arbitrary parameter $\frs$ is performed in Sec.~\ref{S:mesofluct-rs}.

\end{itemize}
The announced program will be realized step by step in Secs.~\ref{S:nonpert}, \ref{S:final} and \ref{S:mesofluct-rs}.

\section{Nonperturbative averaging of eight Green functions}
\label{S:nonpert}

\subsection{Pairwise correlator $\corr{ \Delta G_{\ve} \Delta G_{\ve}}$}
\label{paircorr}

We start with discussing the pair correlation function on the left-hand side of Eq.~(\ref{corr}). It is instructive to introduce a small energy mismatch $\omega$ between the energies of two Green functions and to consider a more general correlation function
\be
\label{R-def}
  \int d\br \, d\br' \,
  \corr{ \Delta G_{\ve}(\br,\br) \Delta G_{\ve-\omega}(\br',\br')}
  =
  -4\pi^2\nu^2 R_\gamma(\omega) .
\ee
As usual, the terms $G^\text{R}G^\text{R}$ and $G^\text{A}G^\text{A}$ are averaged trivially, each contributing 1/4 to $R_\gamma(\omega)$. Averaging of the cross term $G^\text{R}G^\text{A}$ is performed with the help of Efetov's zero-dimensional supersymmetric sigma model \cite{Efetov-book}, where in accordance with Eq.~(\ref{width}) one has to introduce a complex frequency
\be
  \Omega = \eps_+ - (\eps-\omega)_-
  =
  \omega + i[\gamma(\ve)+\gamma(\ve-\omega)]/2 .
\ee
Evaluating the standard integrals with the help of the machinery developed in \ref{slowcontrrules}, we obtain
\be
\label{R-general}
  R_\gamma(\omega)
  =
  1-\Re X(-i\pi\Omega/\Delta) ,
\ee
where the function $X(a)$ is given by Eq.~(\ref{X}):
\be
  X(a)=\frac{1-e^{-2a}}{2a^2} .
\ee

In the limit of vanishing width ($\gamma\to0$), $R_0(\omega)$ reproduces the pair correlation
function for the Gaussian unitary ensemble in the random matrix theory \cite{Mehta}:
$
  R_0^\text{RMT}(\omega) = 1-(\sin x/x)^2+\pi \delta(x)
$,
where $x=\pi\omega/\mls$.
A finite but small width acts as a regularizer of the $\delta$ function, accounting for the contribution of the same broadened level into the correlation function. In the relevant limit of a resolved discrete spectrum, $(\omega,\gamma)\ll\mls$, $R_\gamma(\omega)$ can be written as
\be
\label{R-delta}
  R_\gamma(\omega)
  \approx
  \Delta \delta_\gamma(\omega) ,
\ee
where $\delta_\gamma(\omega)$ is a Lorentzian approximation of the delta function:
\be
\label{delta_gamma}
  \delta_\gamma(\omega)
  =
  \frac{1}{\pi} \frac{\gamma(\ve)}{\gamma^2(\ve)+\omega^2} .
\ee
It is worth noting that Eq.~(\ref{R-delta}) is non-perturbative. Though it looks just like a one-diffuson contribution, $R_\gamma(\omega) \approx -\Im 1/(\omega+i\gamma)$, it is an asymptotic expansion of the exact non-perturbative expression (\ref{R-general}).

The left-hand side of Eq.~(\ref{corr}) can be expressed with the help of Eq.~(\ref{R-def}), with $R_\gamma(0)=\Delta/\pi\gamma(\ve)$.

\subsection{Supersymmetric sigma model for eight Green functions (the case $\frs=0$)}
\label{SS:SUSY}

The objects $X_4X_4'$, $X_4Y_4'$ and $Y_4Y_4'$
contain the products of eight $\Delta G = G^\text{R}-G^\text{A}$ which should be averaged over disorder.
To introduce the method we start with
the simplest case of $\frs\to0$. Generalization to the case of an arbitrary $\frs$ will be performed in Sec.~\ref{S:mesofluct-rs}. To be specific, we focus on calculating $\corr{X_4X_4'}$ (disorder averaging of other products is performed analogously and the result is presented in Sec.~\ref{Seconddiagram}).
As we will see in Secs.~\ref{slowaveraging} and \ref{qual}, only terms with the equal number of retarded and advanced Green functions make the leading contribution to mesoscopic fluctuations.
Consider, e.~g., a particular choice of $G^\text{R}$ and $G^\text{A}$ from $X_4X_4'$ and calculate
\be
K=\langle G^\text{R}_{\ve}(\br_2,\br_1) G^\text{A}_{\ve}(\br_6,\br_5) G^\text{R}_{\ve-\omega_2}(\br_5,\br_6) G^\text{A}_{\ve-\omega_1}(\br_1,\br_2)
G^\text{R}_{\ve_1-\omega_1}(\br_4,\br_3)
G^\text{A}_{\ve_2-\omega_2}(\br_8,\br_7)  G^\text{R}_{\ve_2}(\br_7,\br_8) G^\text{A}_{\ve_1}(\br_3,\br_4) \rangle .
\label{GGGG}
\ee
Disorder averaging of other relevant terms from $X_4X_4'$ [listed in Eq.~(\ref{RA-4ways})] can be performed analogously (see Secs.~\ref{SS:X8a} and \ref{SS:X8b}).

In order to calculate $K$ we follow the standard line of Efetov's supersymmetric sigma model \cite{Efetov-book}, generalizing it to the case of eight Green functions. We group the Green functions into four RA pairs in the sequence they appear in Eq.~(\ref{GGGG}):
\be
\label{1234}
  \begin{tabular}{|c|cccc|}
  \hline
      & 1 & 2 & 3 & 4 \\
  \hline
    R & $\ve$ & $\ve-\omega_2$ & $\ve_1-\omega_1$ & $\ve_2$ \\
  \hline
    A & $\ve$ & $\ve-\omega_1$ & $\ve_2-\omega_2$ & $\ve_1$ \\
  \hline
  \end{tabular}
\ee
thereby introducing a new space of pairs, that will be referred to as 1234.
The way this space is introduced is consistent with the pairing (a) in Eq.~(\ref{energy-matching}) shown in Fig.~\ref{F:AGKL31}(a). The leading contribution to mesoscopic fluctuations from this pairing will come from configurations when the energies in each pair nearly coincide, corresponding to $Q$ matrices which are block-diagonal in the space 1234 [see Eq.~(\ref{Q-blocks}) below].
The other relevant contribution originates from the pairing (b) in Eq.~(\ref{energy-matching}) shown in Fig.~\ref{F:AGKL31}(b).
In principle, it can be also handled using the 1234 structure of Eq.~(\ref{1234}), however the corresponding $Q$ matrices will have a cumbersome structure in this basis.
Therefore in the study of the pairing (b) in Sec.~\ref{SS:X8b} we will introduce 1234 space in a different way consistent with that pairing.

The resulting sigma model is formulated in terms of the $16\times 16$ supermatrix field $Q(\br)$ acting in the tensor product of the spaces $\FB\otimes\RA\otimes 1234$, where FB stands for the superspace. The matrix $Q$ can be written as $Q=T^{-1}\Lambda T$, where $T$ spans the supersymmetric coset
$\text{U}(8|4,4)/\text{U}(4|4)\times\text{U}(4|4)$, and $\Lambda=\sigma_3^\text{RA}$. The sigma-model action is given by (we adopt the fermion-dominated notation of Ref.~\cite{Efetov-book})
\be
  S[Q]
  =
  \frac{\pi\nu}4\int d\br
  \str \bigl[ D\left( \nabla Q(\br)\right)^2+4i\hat E Q (\br) \bigr],
\label{action}
\ee
where $\hat E$ is the diagonal matrix made of the energy arguments [with widths, according to Eq.~(\ref{width})] of the corresponding Green functions:
\be
\label{E8-def}
  \hat E
  =
  \mathop{\rm diag}
\bigl\{
\ve_+,
(\ve-\omega_2)_+, (\ve_1-\omega_1)_+, (\ve_2)_+,
\ve_-,
(\ve-\omega_1)_-, (\ve_2-\omega_2)_-,(\ve_1)_-
\bigr\}\otimes 1_{\FB}.
\ee

In the sigma-model language, the average product of eight Green functions in Eq.~(\ref{GGGG}) transforms to
\be
  K=(\pi \nu)^8\int Q^{\AR}_{21}(\br_1)Q^{\RA}_{12}(\br_2)Q^{\RA}_{21}(\br_5)Q^{\AR}_{12}(\br_6) Q^{\AR}_{43}(\br_3)Q^{\RA}_{34}(\br_4)Q^{\RA}_{43}(\br_7)Q^{\AR}_{34}(\br_8) e^{-S[Q]}DQ,
\label{Qcorr}
\ee
where all the elements of $Q$ matrices in the preexponent are taken from the BB sector (omitted for brevity). Equation (\ref{Qcorr}) holds only in the limit $\frs\to0$, when the interaction range is much larger than the Fermi wave length, and additional terms in the preexponent are suppressed by Friedel oscillations. In the general case discussed in Sec.~\ref{S:mesofluct-rs}, Eq.~(\ref{Qcorr}) should be replaced by Eq.~(\ref{Qcorr-rs}).

\subsection{Integration over fast modes}
\label{fastaveraging}

Besides eight Green functions, the block $X_4X_4'$ given by Eq.~(\ref{X8}) contains four SSI lines, each carrying a non-zero momentum. Therefore in calculating $K$ in Eq.~(\ref{Qcorr}) we should (i) allow fast ($q\neq0$) diffuson modes to make $\corr{X_4X_4'}$ nonzero, (ii) minimize their number as every fast mode brings a small factor of $\Delta/\ETh$, and (iii) be able to handle the resulting zero-dimensional integral non-perturbatively in order to resolve discrete levels. This task can be accomplished following the strategy of Ref.~\cite{KM}, where non-universal corrections to the random-matrix level statistics were calculated beyond the zero-dimensional limit.
For this purpose we write
\be
  Q(\br) = T^{-1} Q'(\br) T,
\label{param}
\ee
where the matrix $T$ is spatially uniform, and $Q'(\br)$ describes all fast modes with non-zero momenta.
Since the latter are to be accounted perturbatively, we expand $Q'(\br)$ near the origin:
\be
  Q'(\br)=\Lambda\left[1+W(\br)+W^2(\br)/2+\ldots\right], \qquad \{W(\br),\Lambda\}=0.
\label{Q'-W}
\ee

To get the leading contribution to $\corr{X_4X_4'}$ we extract one fast $W$ from each of the eight $Q$ matrices in the preexponent of Eq.~(\ref{Qcorr}), and average it with the Gaussian action
\be
  S^{(2)}[W] = - \frac{\pi\nu D}4 \int d\br\, \str[\nabla W(\br)]^2 ,
\ee
coming from the gradient term of Eq.~(\ref{action}).
Using Wick's theorem, the correlator of eight $W$ fields can be expressed as the sum of all possible products of four pairwise correlators. The latter can be easily calculated with the help of the following contraction rule valid for arbitrary matrices $P$ and $R$:
\be
  \langle \str PW(\br) \str RW(\br') \rangle_W
  =
  D(\br,\br')\str(P\Lambda R\Lambda -PR),
\qquad
  D(\br,\br')=\frac{\corr{\br|(-\nabla^2)^{-1}|\br'}}{\pi\nu D} .
\label{contruction rules}
\ee
According to Wick's theorem, averaging over fast modes generates $7!!=105$ different terms in the expression for $K$.
Not all of them are equally important. To pick up the relevant terms one should understand how to perform further integration over the zero mode. This procedure will be discussed in Sec.~\ref{slowaveraging}, and in Sec.~\ref{SS:further} we will proceed with the derivation based on Eq.~(\ref{contruction rules}).

\subsection{Block structure of the zero-dimensional sigma model and energy pairs}
\label{slowaveraging}

Having integrated out fast diffusive modes, we end up with the effective zero-dimensional sigma model for the $16 \times 16$ supermatrix $Q=T^{-1}\Lambda T$ with the action
\be
\label{action0}
  S_0[Q]=\frac{\pi i }{\Delta}\str \hat E Q.
\ee
Owing to a large size of the matrix $Q$, this is still a complicated theory. Fortunately, we do not need its exact solution since, as discussed in Sec.~\ref{S:MF}, the most singular contribution to $\ccorr{\gamma^2(\eps,T)}$ comes from pairwise coinciding energies, see Eq.~(\ref{energy-matching}). Since we are interested in the limit $(\eps,T)\gg\mls$, energy difference between pairs is typically large, $|\eps_i-\eps_j|\gg\mls$, and hence correlations between different pairs can be neglected (configurations with $|\eps_i-\eps_j|\lesssim\mls$ which require non-perturbative treatment exist but their weight is small). On the other hand, energies from the same pair match with the accuracy of $\gamma$ and should be treated non-perturbatively like in Sec.~\ref{paircorr}. In the language of the zero-dimensional sigma model, large energy difference between pairs suppresses degrees of freedom in the supermatrix $Q$ which are off-diagonal in the space of pairs. As a result, $Q$ becomes block-diagonal in the space of pairs, and the full theory splits into a product of four standard Efetov's supermatrix sigma models for $\corr{G^\text{R}G^\text{A}}$, see Eq.~(\ref{S-splitting}) below.

\begin{figure}
\centerline{\includegraphics[width=1.0\textwidth]{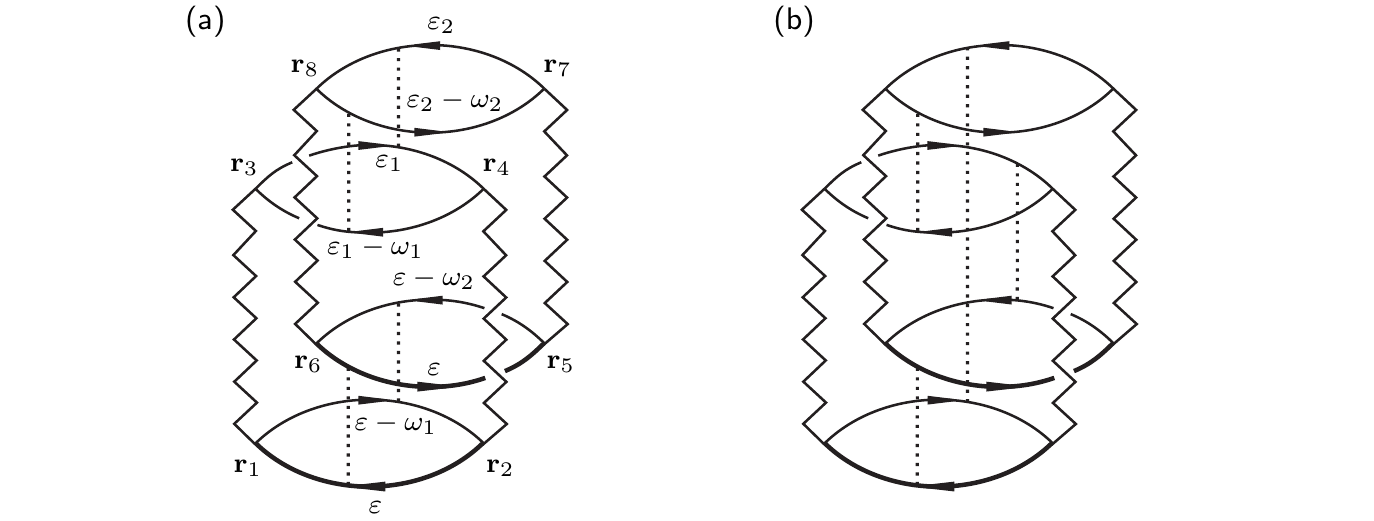}}
\caption{
Two types of arranging eight Green functions in
the product
$X_4X_4'$ [Eq.~(\ref{X8})] into four pairs that produce the most singular contributions $\corr{X_4X_4'}_\text{sing}^{(a)}$ and $\corr{X_4X_4'}_\text{sing}^{(b)}$ in Eq.~(\ref{X8-sing-reg}).
Green functions from the same pair are marked by a dotted line. The pairings (a) and (b) correspond to the two ways to match energies in Eq.~(\ref{energy-matching}). All energy and coordinate indices in (b) are the same as in (a).}
\label{F:AGKL31}
\end{figure}

The next step is to understand what are the possible ways to group eight Green functions into four pairs to maximize their contribution to fluctuations of $\gamma(\eps,T)$. First of all, it is clear that the Green functions with the energies $\ve$ always need to be in a pair, as they immediately produce the smeared delta function of zero argument, $\delta_\gamma(0)=1/\pi\gamma$ [cf.\ Eq.~(\ref{delta_gamma})]. It is this large factor that compensates the analogous contribution from $\corr{G^\text{R}_\eps G^\text{A}_\eps}$ on the left-hand side of Eq.~(\ref{corr}). Our aim then is to identify those pairings that may introduce an additional large factor of $\delta_\gamma(0)$ after integration over all intermediate energies. Such a situation can be realized only if coincidence of the energies in two pairs automatically implies coincidence of the energies in the third pair. That can be achieved only in two physically relevant cases, (a) and (b), listed in Eq.~(\ref{energy-matching}) and shown diagrammatically in Fig.~\ref{F:AGKL31}
(there exist three other unphysical pairings which should be discarded as demonstrated in \ref{A:spurious}). We emphasize that possible choices of pairing are dictated by the energy arguments of electron Green functions and are not specific for the particular arrangement of $G^\text{R}$ and $G^\text{A}$. The only  natural requirement is that each pair contains one retarded and one advanced Green function (otherwise, there is no correlations within a pair at all).

According to this general logic, each diagram for $\corr{X_4X_4'}$, $\corr{X_4Y_4'}$ and $\corr{Y_4Y_4'}$ can be written in the form
\be
  \corr{A} = \corr{A}_\text{sing}^{(a)} + \corr{A}_\text{sing}^{(b)} + \corr{A}_\text{reg} ,
\label{X8-sing-reg}
\ee
where $\corr{\dots}_\text{sing}^{(a)}$ and $\corr{\dots}_\text{sing}^{(b)}$ denote singular (in the limit $\gamma\to0$) contributions from the pairings (a) and (b).
We proceed with calculating these singular contributions for the term $K$ [a particular element of $\corr{X_4X_4'}$ defined in Eq.~(\ref{GGGG})] from the zero-dimensional sigma model (\ref{action0}).
The introduction of the 1234 space in Eq.~(\ref{1234}) is consistent with the pairing (a): in the 1234 basis the corresponding matrix $Q$ becomes diagonal. On the contrary, the pairing (b) is described by non-diagonal matrices in the 1234 space. To calculate $K_\text{sing}^{(b)}$ we find it more convenient to change the basis and introduce a new $1234'$ space consistent with the pairing (b), in which the matrix $Q$ is diagonal.
Then the calculation of $K_\text{sing}^{(b)}$ becomes completely analogous to the calculation of $K_\text{sing}^{(a)}$.
To illustrate the technique, we focus on the contribution of $K$ due to the pairing (a) [the contribution of the pairing (b) is discussed in Sec.~\ref{SS:X8b}].

\subsection{Singular contribution of the pairing (a)}
\label{SS:further}

Now we are ready to proceed with the calculation started in Sec.~\ref{fastaveraging}.
In order to apply the contraction rules (\ref{contruction rules}) for averaging the preexponent in Eq.~(\ref{Qcorr}) over fast modes, it is convenient to introduce projectors onto different sectors of the $Q$ manifold.
By definition, we write $Q^{ab}_{ij}(\br)=\str P^{ab}_{ij}Q(\br)$, where $a$ and $b$ refer to the RA space, whereas $i$ and $j$ refer to the 1234 space (BB sector is also implied). For example, the projector $P^{\AR}_{21}$ is given by
\be
  P^{\AR}_{21} = \begin{pmatrix} 0&P^{\AR}&0&0\\0&0&0&0\\0&0&0&0\\0&0&0&0 \end{pmatrix}_{1234},
\quad
  P^{\AR} = \begin{pmatrix} 0&P_\text{BB}\\0&0 \end{pmatrix}_{\RA},
\quad
  P_\text{BB} = \begin{pmatrix} 0&0\\0&1 \end{pmatrix}_{\FB}.
\ee

As we discussed in Sec.~\ref{slowaveraging}, the singular contribution $K_\text{sing}^{(a)}$ originates from the matrices $Q$ which are diagonal in $1234$ space. Such a structure of the $Q$ matrix guarantees that only four out of $7!!=105$ terms appearing after averaging over fast modes [see Eq.~(\ref{param})] turn out to be non-zero:
\begin{align}
  K_\text{sing}^{(a)}
  =
  (\pi\nu)^8
  \bigl\langle\bigl\{
    D(\br_1,\br_2) & D(\br_5,\br_6)\str \left(P^{\AR}_{21}QP^{\RA}_{12}Q-P^{\AR}_{21} P^{\RA}_{12}\right) \str \left(P^{\RA}_{21}QP^{\AR}_{12}Q-P^{\RA}_{21} P^{\AR}_{12}\right)
\nonumber
\\{}
  + D(\br_1,\br_6) & D(\br_2,\br_5) \str \left(P^{\AR}_{21}QP^{\AR}_{12}Q-P^{\AR}_{21} P^{\AR}_{12}\right) \str \left(P^{\RA}_{21}QP^{\RA}_{12}Q- P^{\RA}_{21} P^{\RA}_{12}\right)\bigr\}
\nonumber
\\{}
\times \bigl\{ D(\br_3,\br_4) & D(\br_7,\br_8)\str \left(P^{\AR}_{43}QP^{\RA}_{34}Q-P^{\AR}_{43} P^{\RA}_{34}\right) \str \left(P^{\RA}_{43}QP^{\AR}_{34}Q-P^{\RA}_{43} P^{\AR}_{34}\right)
\nonumber
\\{}
+ D(\br_3,\br_8) & D(\br_4,\br_7) \str \left(P^{\AR}_{43}QP^{\AR}_{34}Q-P^{\AR}_{43} P^{\AR}_{34}\right) \str \left(P^{\RA}_{43}QP^{\RA}_{34}Q- P^{\RA}_{43} P^{\RA}_{34}\right)\bigr\}\bigr\rangle
,
\label{Ka}
\end{align}
where $\langle \ldots \rangle$ denotes averaging over the $16\times16$ zero-dimensional sigma-model manifold with the action (\ref{action0}).

Since $Q$ is diagonal in the 1234 space, it can be represented as
\be
\label{Q-blocks}
  Q = \mathop{\rm diag} \{Q_1, Q_2, Q_3, Q_4\} ,
\ee
where $Q_j$ are independent $4\times4$ supermatrices in the $\RA\otimes\FB$ space spanning the standard supermanifold of the sigma model for $\corr{G^\text{R}G^\text{A}}$.
The action (\ref{action0}) then splits into a sum of four separate Efetov's sigma-model actions:
\be
\label{S-splitting}
  S_0[Q]
  =
  \sum_{j=1}^4 S_{0j}[Q_j]
  =
  \frac{\pi i}{\Delta} \sum_{j=1}^4 \str \hat E_j Q_j,
\ee
where the elements of the diagonal matrix $\hat E_j$ are taken from the corresponding sector of $\hat E$ defined in Eq.~(\ref{E8-def}). Now averaging over all four $Q_j$ is performed independently and can be done exactly in the standard way \cite{Efetov-book}. To handle the supertrace structure in Eq.~(\ref{Ka}), it is convenient to use the zero-dimensional contraction rules (\ref{contr-rules-0D}) derived in \ref{slowcontrrules}. The result can be expressed in a compact form in terms of the quantities
\be
  X_j \equiv X\left(-i\pi\Omega_j/\Delta \right) ,
\qquad
  Z_j \equiv Z\left(-i\pi\Omega_j/\Delta \right) , \label{XjZj}
\ee
where the functions $X(a)$ and $Y(a)$ are given by Eqs.~(\ref{X}) and (\ref{Z}):
\be
\label{XZ}
  X(a)=\frac{1-e^{-2a}}{2a^2} ,
\qquad
  Z(a)=\frac 1a ,
\ee
and $\Omega_j$ is the energy difference between the arguments of $G^\text{R}$ and $G^\text{A}$ in the corresponding pair [cf.\ Eq.~(\ref{1234})]:
\begin{gather}
\label{Omega1234}
  \Omega_1 = \ve_+ - \ve_- ,
  \!\!\qquad
  \Omega_2 = (\ve-\omega_2)_+ - (\ve-\omega_1)_- ,
  \!\!\qquad
  \Omega_3 = (\ve_1-\omega_1)_+ - (\ve_2-\omega_2)_-  ,
  \!\!\qquad
  \Omega_4 = (\ve_2)_+ - (\ve_1)_- .
\end{gather}

To demonstrate the technique, consider for example one of the four terms in Eq.~(\ref{Ka}):
\be
  K_\text{sing}^{(a)}
  =
  (\pi\nu)^8 D(\br_1,\br_2)D(\br_5,\br_6) D(\br_3,\br_4)D(\br_7,\br_8) L_1 + \dots
\label{Ka2}
\ee
Substituting $Q$ in the form (\ref{Q-blocks}) and tracing over the 1234 space we obtain
\begin{align}
  L_1
  =
  \bigl\langle & \str (P^{\AR}Q_2P^{\RA}Q_1-P^{\AR}_{21} P^{\RA}_{12}) \str (P^{\RA}Q_2P^{\AR}Q_1-P^{\RA}_{21} P^{\AR}_{12})
\nonumber
\\{} \times
  & \str (P^{\AR}Q_4P^{\RA}Q_3-P^{\AR}_{43} P^{\RA}_{34}) \str (P^{\RA}Q_4P^{\AR}Q_3-P^{\RA}_{43} P^{\AR}_{34})\bigr\rangle.
\label{K11}
\end{align}
Sequentially applying the contraction rules (\ref{contr-rules-0D}) for averaging over all $Q_j$, we arrive at the exact non-perturbative expression
\be
  L_1=(4+2X_1+2X_2+4X_1X_2)(4+2X_3+2X_4+4X_3X_4) .
\ee
Contributions of the three other terms in Eq.~(\ref{Ka2}) are calculated analogously, and we obtain finally the singular part of $K$ due to pairing (a):
\begin{align}
  K_\text{sing}^{(a)}
  =(\pi\nu)^8
  & \left\{D(\br_1,\br_2)D(\br_5,\br_6)(4+2X_1+2X_2+4X_1X_2)+ 4D(\br_1,\br_6)D(\br_2,\br_5)Z_1Z_2\right\}
\nonumber
\\{}
  \times
  & \left\{ D(\br_3,\br_4)D(\br_7,\br_8)(4+2X_3+2X_4+4X_3X_4)+4D(\br_3,\br_8)D(\br_4,\br_7)Z_3Z_4\right\}.
\label{K1}
\end{align}

Though the singular contribution due to the energy pairing (a) may originate only from Eq.~(\ref{K1}), not all terms in this equation do actually produce the singular contribution. We analyze this expression and perform the final step of calculation in the next Section.

\section{Mesoscopic fluctuations at $\frs=0$}
\label{S:final}

\subsection{General recipe}
\label{qual}

The most singular contribution from Eq.~(\ref{K1}) originates from the terms which contain the maximal number (four) of $X_j$ and $Z_j$. We have already explained that this choice is dictated by the necessity to obtain two smeared delta-functions (\ref{delta_gamma}) with zero argument, contributing a large factor $\Delta/\gamma$ each.
This is also the reason why we have to choose each pair in the $1234$ space to consist of one $G^R$ and one $G^\text{A}$: large factors $X_j$ and $Z_j$ appear as a result of averaging $\langle G^\text{R} G^\text{A} \rangle$ over the zero-dimensional diffusive model.

In calculating the energy integrals in Eq.~(\ref{corr}), the important contribution comes from the vicinity of the poles of $X_j$ and $Z_j$. This observation allows us to work in the limit $\Omega_j \sim \gamma \ll\mls$ [with $\Omega_j$ introduced in Eq.~(\ref{Omega1234}) being the energy mismatch within a pair] and use the leading-order asymptotics of Eqs.~(\ref{XZ}):
\be
  X_j \approx Z_j = \frac{i\Delta}{\pi\Omega_j} .
\ee
The chosen sequence of $G^\text{R}$ and $G^\text{A}$ in the correlator $K$ [Eq.~(\ref{GGGG})] also ensures that the product $1/\Omega_2\Omega_3\Omega_4$ has poles both in the upper and lower half-planes of the energy variables $\ve_1$, $\ve_2$, $\omega_1$ and $\omega_2$. Hence integration over intermediate energies does not vanish and indeed produces the desired delta-function-like contribution.

Now the recipe for extracting the leading singular term from Eq.~(\ref{K1}) and  similar expressions for other diagrams can be formulated as follows.
One should substitute all $X_j$ by $Z_j$ and take the term proportional to ${\cal Z} = Z_1Z_2Z_3Z_4$.
Then the coefficient ${\cal Z}$ should be replaced by
\be
  {\cal Z}
  \longrightarrow
  \frac{4\Delta^4}{\pi^2 \gamma(\ve)}
  \frac{\delta(\ve_1-\ve_2)\delta(\omega_1-\omega_2)}{\gamma(\ve-\omega_1)+\gamma(\ve_1)+\gamma(\ve_1-\omega_1)} ,
\label{XXX}
\ee
where the factor $\Delta/\pi\gamma(\ve)$ originates from $Z_1$ and will be canceled by a similar factor from the left-hand side of Eq.~(\ref{corr}) (see Sec.~\ref{paircorr}), while the coefficient in front of the delta functions can be easily obtained by calculating, e.g., $\int Z_2Z_3Z_4\,d\ve_2d\omega_2$.
The last step is to integrate four diffuson propagators in Eq.~(\ref{K1}) over $\br_i$ [see Eq.~(\ref{corr})].
In doing that, $\delta_\kappa(\br-\br')$ in the expressions for
$X_4X_4'$, $X_4Y_4'$ and $Y_4Y_4'$
can be considered as the usual zero-range delta function.
Then depending on the term under consideration, diffusons combine either into $(\tr D^2)^2$ or  $\tr D^4$. These traces evaluated in the Fourier space produce either $1/E_2^4$ or $1/E_4^4$, where $E_n\sim\ETh$ are defined in Eq.~(\ref{En}).
As a result, the contribution to mesoscopic fluctuations of the inelastic rate can be written in the form
\be
  \ccorr{\gamma^2(\ve,T)}_{A}^{(i)}
  =
  \frac{{\lambda^4}\Delta^5}{4\pi^3}
  c_{A}^{(i)}
  \int d\ve_1 d\omega_1
  \frac{(\FF_{\ve_1}-\FF_{\ve_1-\omega_1})^2(\BB_{\omega_1}+\FF_{\ve-\omega_1})^2}
  {\gamma_0(\ve-\omega_1,T)+\gamma_0(\ve_1,T)+\gamma_0(\ve_1-\omega_1,T)}
  ,
\label{gamma-c}
\ee
where $A$ is a particular diagram or a set of diagrams considered, and $i$ labels the type of the energy pairing (a or b).

\subsection{Contribution from $X_4X_4'$ due to the energy pairing (a)}
\label{SS:X8a}

Applying this general recipe to expression (\ref{K1}) for $K_\text{sing}^{(a)}$, we first reduce it to the form
\be
  K_\text{sing}^{(a)}
  =16(\pi\nu)^8 \, {\cal Z}
  \left\{D(\br_1,\br_2)D(\br_5,\br_6) + D(\br_1,\br_6)D(\br_2,\br_5)\right\}
  \left\{ D(\br_3,\br_4)D(\br_7,\br_8) + D(\br_3,\br_8)D(\br_4,\br_7)\right\} ,
\label{K1-Z}
\ee
and then, tracing the diffuson propagators, obtain the coefficient $c_{K}^{(a)}$ in Eq.~(\ref{gamma-c}):
\be
  c_{K}^{(a)}
  =
  \frac{1}{2\ns^2}\left(\frac 1{E_2^4}+\frac 1{E_4^4}\right) .
\label{gamma1'}
\ee

To complete the analysis of the contribution from $X_4X_4'$ due type-(a) energy pairing, one has to consider other arrangements of $G^\text{R}$ and $G^\text{A}$. The requirement of having one $G^\text{R}$ and one $G^\text{A}$ in each pair limits the number of various possibilities to $2^4=16$. However, not all of them should be taken into account. Consider, for example, the correlator
$$
K'=\langle G^\text{R}_{\ve}(\br_2,\br_1) G^\text{A}_{\ve}(\br_6,\br_5)  G^\text{R}_{\ve-\omega_2}(\br_5,\br_6) G^\text{A}_{\ve-\omega_1}(\br_1,\br_2)
G^\text{R}_{\ve_1-\omega_1}(\br_4,\br_3) G^\text{A}_{\ve_2-\omega_2}(\br_8,\br_7) G^\text{A}_{\ve_2}(\br_7,\br_8) G^\text{R}_{\ve_1}(\br_3,\br_4)\rangle ,
$$
which differs from $K$ [Eq.~(\ref{GGGG})] by changing R${}\leftrightarrow{}$A in the last two Green functions.
The expression analogous to Eq.~(\ref{K1}) will contain  $X_4^*$ instead of $X_4$ and $Z_4^*$ instead of $Z_4$, which renders the poles of both $X_3X_4^*$ and $Z_3Z_4^*$  lying in the upper half-plane of $\ve_2$. Therefore, the contribution of this term is non-singular and should be disregarded.
A simple analysis demonstrates that there are only four  possibilities to arrange $G^\text{R}$ and $G^\text{A}$
in $X_4X_4'$:
\be
\label{RA-4ways}
  \text{RARARARA},
\quad
  \text{ARRARARA},
\quad
  \text{RAARARAR},
\quad
  \text{ARARARAR},
\ee
where all energy and coordinate indices follow Eq.~(\ref{GGGG}). Each choice gives the same contribution given by Eq.~(\ref{gamma1'}). As a result, the total contribution of the term $X_4X_4'$ due to type-(a) energy pairing (shown in Fig.~\ref{F:AGKL31}a) is described by the coefficient
\be
  c_{XX}^{(a)}
  =
  \frac{2}{\ns^2}\left(\frac 1{E_2^4}+\frac 1{E_4^4}\right) .
\label{gamma11}
\ee

\subsection{Contribution from $X_4X_4'$ due to the energy pairing (b)}
\label{SS:X8b}

Following the same line we can analyze the singular part of $K$ due to pairing (b) shown in Fig.~\ref{F:AGKL31}(b). Instead of Eq.~(\ref{K1-Z}) we obtain:
\be
  K_\text{sing}^{(b)}
  =
  16 (\pi\nu)^8 \, {\cal Z} \,
  D(\br_1,\br_2) D(\br_3,\br_4)
  D(\br_5,\br_6) D(\br_7,\br_8) .
\label{K2-Z}
\ee
Following the procedure described in Sec.~\ref{qual} and utilizing the same four possibilities (\ref{RA-4ways}) to arrange $G^\text{R}$ and $G^\text{A}$, we arrive at
\be
  c_{XX}^{(b)}
  =
  \frac{1}{\ns^2E_2^4} .
\label{gamma12}
\ee

\subsection{Contributions from $X_4Y_4'$ and $Y_4Y_4'$}
\label{Seconddiagram}

It is left to discuss the other two terms, $X_4Y_4'$ and $Y_4Y_4'$, in Eq.~(\ref{corr}). Following the same steps we obtain that the contributions of the cross term $\corr{X_4Y_4'}$ from both the pairings (a) and (b) vanish after integration over fast diffusive modes due to the diagonal structure of the $Q$ matrix in the 1234 space, while in calculating $\corr{Y_4Y_4'}$ only the pairing (b) should be taken into account for the same reason.

Consider, e.~g., a particular realization of $G^\text{R}$ and $G^\text{A}$ from $\corr{Y_4Y_4'}$ [compare with $K$ defined in Eq.~(\ref{GGGG})]:%
\be
\tilde K=\langle G^\text{R}_{\ve}(\br_4,\br_1) G^\text{A}_{\ve_1}(\br_1,\br_2) G^\text{R}_{\ve_1-\omega_1}(\br_2,\br_3) G^\text{A}_{\ve-\omega_1}(\br_3,\br_4) G^\text{A}_{\ve}(\br_8,\br_5) G^\text{R}_{\ve_2}(\br_5,\br_6) G^\text{A}_{\ve_2-\omega_2}(\br_6,\br_7) G^\text{R}_{\ve-\omega_2}(\br_7,\br_8) \rangle .
\ee
Its singular contribution due to the pairing (b) is given by
\be
  \corr{\tilde K}_\text{sing}^{(b)}
  =
  16 (\pi\nu)^8 \, {\cal Z} \,
  D(\br_1,\br_2) D(\br_3,\br_4)
  D(\br_5,\br_6) D(\br_7,\br_8) ,
\ee
Tracing the diffusons and taking into account four possibilities (\ref{RA-4ways}), we get
\be
  c_{YY}^{\text{(b)}}
  =
  \frac{1}{\ns^4E_2^4} .
\label{gamma2}
\ee

Finally, adding the contributions (\ref{gamma11}), (\ref{gamma12}) and (\ref{gamma2}) we obtain the final expression for mesoscopic fluctuations of the inelastic rate in the limit $\frs\to0$:
\be
  \ccorr{\gamma^2(\ve,T)}
  =
  \frac{\lambda^4\Delta^5}{4\pi^3\ns^4}
  \left(\frac{3\ns^2+1}{E_2^4}+\frac{2\ns^2}{E_4^4}\right)
  \int d\ve_1 d\omega_1
  \frac{(\FF_{\ve_1}-\FF_{\ve_1-\omega_1})^2(\BB_{\omega_1}+\FF_{\ve-\omega_1})^2}
  {\gamma_0(\ve-\omega_1,T)+\gamma_0(\ve_1,T)+\gamma_0(\ve_1-\omega_1,T)}
  .
\label{gamma-total-f0}
\ee

\section{Mesoscopic fluctuations at arbitrary $\frs$}
\label{S:mesofluct-rs}

\subsection{Account for a finite $\frs$ in the sigma-model language}

In this Section we generalize the result (\ref{gamma-total-f0}) to the case of an arbitrary interaction parameter $\frs$ and derive the general expression (\ref{subfinal}).
First, we explain the main idea with an example of the type-(a) energy pairing in the diagram for $X_4X_4'$, and then apply it to the other diagrams ($X_4Y_4'$ and $Y_4Y_4'$) and energy pairing (b).

When the range of the SSI (\ref{V(r)}) characterized by the function $\delta_\kappa(\br)$ becomes comparable to the Fermi wave length (i.~e., $\frs$ becomes non-negligible), a simple sigma-model expression (\ref{Qcorr}) for the correlator $K$ breaks down. To modify it one has to turn back to the intermediate step in the derivation of the sigma model, prior to the final integration over the 16-component supervector $\psi$ used to represent Green functions in the functional form.
At this stage, the correlator $K$ [Eq.~(\ref{GGGG})] is written as
\begin{align}
  K
  =
  \int
  \langle
& \psi_\text{R1}(\br_2) \psi^*_\text{R1}(\br_1)
  \psi_\text{A1}(\br_6) \psi^*_\text{A1}(\br_5)
  \psi_\text{R2}(\br_5) \psi^*_\text{R2}(\br_6)
  \psi_\text{A2}(\br_1) \psi^*_\text{A2}(\br_2)
\nonumber
\\{} \times{}
& \psi_\text{R3}(\br_4) \psi^*_\text{R3}(\br_3)
  \psi_\text{A3}(\br_8) \psi^*_\text{A4}(\br_7)
  \psi_\text{R4}(\br_7) \psi^*_\text{R4}(\br_8)
  \psi_\text{A4}(\br_3) \psi^*_\text{A4}(\br_4)
  \rangle_{\psi}
  \,
  e^{-S[Q]} DQ ,
\label{psi16}
\end{align}
where the bosonic components of the superfield $\psi$ are implied.
Averaging over $\psi$ should be performed with the help of Wick's theorem with the correlation function
\be
  \corr{\psi(\br)\psi^*(\br')} = -i g(\br,\br') \Lambda ,
\ee
where $g$ is the supermatrix Green function defined as
\be
  g(\br,\br')
  =
  \langle
    \br | ( \hat E - H_0 + iQ/2\tau )^{-1} | \br'
  \rangle .
\ee

The result of the $\psi$-averaging is sensitive to the distance between the points $\br_i$ controlled by the spatial range of the SSI propagators through the factors $\delta_\kappa(\br_i-\br_j)$ in Eq.~(\ref{X8}).
Since the interaction is assumed to be short-range on the scale of the system size $L$, only correlations between $\psi$'s coupled to the same SSI propagator should be taken into account. Therefore the $\psi$-averaging in Eq.~(\ref{psi16}) factorizes into four contributions corresponding to four SSI propagators in Eq.~(\ref{X8}):
\be
\label{K-KKKK}
  K
  =
  \int K^{(1,3)} K^{(2,4)} K^{(5,7)} K^{(6,8)} e^{-S[Q]} DQ .
\ee
Consider for example the group of four $\psi$'s coupled by the interaction with $\delta_\kappa(\br_1-\br_3)$:
\be
  K^{(1,3)}
  =
  \langle
  \psi^*_\text{R1}(\br_1)
  \psi_\text{A2}(\br_1)
  \psi^*_\text{R3}(\br_3)
  \psi_\text{A4}(\br_3)
  \rangle_{\psi}
.
\label{psi4}
\ee
Application of Wick's theorem generates two terms:
\be
  K^{(1,3)}
  =
  -
  g_\text{A2,R1}(\br_1,\br_1)
  g_\text{A4,R3}(\br_3,\br_3)
  -
  g_\text{A2,R3}(\br_1,\br_3)
  g_\text{A4,R1}(\br_3,\br_1)
.
\label{K13-gg}
\ee

The first (diagonal) term in Eq.~(\ref{K13-gg}) reduces to the $Q$ matrix due to the self-consistency equation $g(\br,\br)=-i\pi\nu Q(\br)$ \cite{Efetov-book}, and then one immediately recovers Eq.~(\ref{Qcorr}) used previously in the analysis of the $\frs=0$ case. In contrast, the second (off-diagonal) term in Eq.~(\ref{K13-gg}) cannot be so easily expressed in terms of the $Q$ matrix. But here we can use the knowledge that at the later stage, in the process of averaging over fast modes (see Sec.~\ref{fastaveraging}), each $Q$ will be expanded to the first power of $W$ [see Eqs.~(\ref{param}), (\ref{Q'-W})]. The linear-in-$W$ contribution to $g$ is given by:
\be
\label{dg1}
  \delta g(\br_1,\br_3)
  =
  - \frac{i}{2\tau}
  \int d\br'
  \langle
    \br_1 | ( - H_0 + iQ/2\tau )^{-1} | \br'
  \rangle
  \,
  T^{-1} \Lambda W(\br') T
  \,
  \langle
    \br' | ( - H_0 + iQ/2\tau )^{-1} | \br_3
  \rangle ,
\ee
where we write $Q=T^{-1}\Lambda T$, and neglect the term $\hat E$ as we are working in the diffusive limit $\max\{\ve, T\} \tau \ll 1$.
Since $W$ anticommutes with $\Lambda$, we can rewrite Eq.~(\ref{dg1}) in the form
\be
\label{dg2}
  \delta g(\br_1,\br_3)
  =
  - \frac{i}{2\tau}
  \int d\br'
  \,
  T^{-1}
  \langle
    \br_1 | ( - H_0 + i\Lambda/2\tau )^{-1} | \br'
  \rangle
  \langle
    \br' | ( - H_0 - i\Lambda/2\tau )^{-1} | \br_3
  \rangle
  \Lambda W
  T
  =
  - \frac{i \delta Q}{2\tau}
  (G^\text{R}G^\text{A})(\br_1-\br_3)
  ,
\ee
where $\delta Q=T^{-1} \Lambda W T$,
and $G^\text{R}G^\text{A}$ is a scalar function given by Eq.~(\ref{GRGA}).

Multiplying the second term of Eq.~(\ref{K13-gg}) by $\delta_\kappa(\br_1-\br_3)$ and integrating over the difference $\br_1-\br_3$, we immediately recover the factor $\frs$ introduced in Eq.~(\ref{f-I}).
Hence, Eq.~(\ref{K13-gg}) can be written as
\be
  K^{(1,3)}
  =
  (\pi\nu)^2
  \left[
  Q^{\AR}_{21}(\br_1) Q^{\AR}_{43}(\br_3)
  + \frs\,
  Q^{\AR}_{23}(\br_1) Q^{\AR}_{41}(\br_3)
  \right] ,
\label{K13-QQ}
\ee
where with our accuracy we do not distinguish between $\br_1$ and $\br_3$.
Generally speaking, Eq.~(\ref{K13-QQ}) is incorrect.
We write it in such a form for brevity, assuming that it will be further subject to the procedure of fast mode extraction described in Sec.~\ref{fastaveraging}.

Performing the same analysis for other $K^{(i,j)}$ from Eq.~(\ref{K-KKKK}), one concludes that for the purpose of calculation of mesoscopic fluctuations of the relaxation rate at $\frs\neq0$, Eq.~(\ref{Qcorr}) should be modified as
\begin{multline}
  K
  =
  (\pi \nu)^8\int
  \left[
  Q^{\AR}_{21}(\br_1) Q^{\AR}_{43}(\br_3)
  + \frs\,
  Q^{\AR}_{23}(\br_1) Q^{\AR}_{41}(\br_3)
  \right]
  \left[
  Q^{\RA}_{12}(\br_2) Q^{\RA}_{34}(\br_4)
  + \frs\,
  Q^{\RA}_{14}(\br_2) Q^{\RA}_{32}(\br_4)
  \right]
\\{}
  \times
  \left[
  Q^{\RA}_{21}(\br_5) Q^{\RA}_{43}(\br_7)
  + \frs\,
  Q^{\RA}_{23}(\br_5) Q^{\RA}_{41}(\br_7)
  \right]
  \left[
  Q^{\AR}_{12}(\br_6) Q^{\AR}_{34}(\br_8)
  + \frs\,
  Q^{\AR}_{14}(\br_6) Q^{\AR}_{32}(\br_8)
  \right]
  e^{-S[Q]}DQ,
\label{Qcorr-rs}
\end{multline}
where the difference between the coordinates within the same bracket can be neglected.
Similar expressions originate in the analysis of other contributions from
$X_4X_4'$, $X_4Y_4'$ and $Y_4Y_4'$.

\subsection{Contributions of different diagrams and energy pairings}

To find mesoscopic fluctuations at a finite $\frs$, one should use Eq.~(\ref{Qcorr-rs}) as a starting point and perform all the steps outlined in Secs.~\ref{S:nonpert} and \ref{S:final}.
This is a routine procedure leading to the following results.
In the case $\frs\neq0$, there are finite contributions from all the terms $X_4X_4'$, $X_4Y_4'$, and $Y_4Y_4'$, and from both energy pairings (a) and (b).
These six contributions are characterized by six coefficients $c_A^{(i)}$ in Eq.~(\ref{gamma-c}). After some algebra, we obtain (the coefficients $c_{XY}^{(i)}$ already contain the factor 2 accounting for the mirror term $Y_4X_4'$):
\begin{gather}
  c_{XX}^{(a)} =
  \frac{2(1+\frs^2)^2}{\ns^2 E_2^4} + \frac{2(1+\frs^4)}{\ns^2 E_4^4} ,
\qquad
  c_{XX}^{(b)} =
  \frac{1+6\frs^2+\frs^4}{\ns^2 E_2^4} + \frac{4\frs^2}{\ns^2 E_4^4} ,
\label{gX}
\\
  c_{XY}^{(a)} = c_{XY}^{(b)} =
  - \frac{8(\frs+\frs^3)}{\ns^3 E_2^4} - \frac{4(\frs+\frs^3)}{\ns^3 E_4^4} ,
\label{gXY}
\\
  c_{YY}^{(a)} = \frac{8 \frs^2}{\ns^4 E_2^4} + \frac{4 \frs^2}{\ns^4 E_4^4} ,
\qquad
  c_{YY}^{(b)} = \frac{1+6\frs^2+\frs^4}{\ns^4 E_2^4} + \frac{4\frs^2}{\ns^4 E_4^4}.
\label{gY}
\end{gather}
Summing the contributions (\ref{gX})--(\ref{gY}), we come to the final result (\ref{subfinal}) valid for an arbitrary $\frs$.

\section{Discussion and conclusion}
\label{Discussion}

In this work we have analyzed the applicability of the Fermi-golden-rule description of the initial stage of quasiparticle decay in diffusive quantum dots in the regime when the single-particle levels are already resolved. Approaching the problem from the high energy/temperature side, where each energy level can be characterized by a Lorentzian width $\gamma_0(\eps,T)$, we have calculated mesoscopic fluctuations of the energy relaxation rate.
The leading contribution to fluctuations comes from the diagrams which describe the square of the same decay process, i.e.\ have the same set of final states.
The resulting expression is non-perturbative in $\gamma_0$, which appears in the denominator of Eq.~(\ref{subfinal}), ensuring the growth of fluctuations with the decrease of the excitation energy and/or temperature.

Quantum relaxation of the initial state $|i\rangle$ can be described by the return probability
\be
  P(t) = \overline{\bigl| \corr{i|e^{-iHt}|i} \bigr|^2} ,
\ee
where $H$ is the Hamiltonian of the interacting quantum dot, and the bar stands for the thermal average. In the semiclassical FGR picture this is just a pure exponential decay:
\be
  P_\text{FGR}(t) = e^{-\gamma(\eps,T)t} ,
\ee
where the rate $\gamma(\eps,T)$ depends on a particular disorder realization.
Its average value is given by Eq.~(\ref{SIAT}), and we focused on its mesoscopic fluctuations.
We found that the FGR description of the initial stage of quasiparicle decay is applicable as long as $\max\{\eps,T\}\gg\EFGR$. In this limit, each level is characterized by a well-defined energy width $\gamma(\eps,T)$, which weakly fluctuates near the FGR mean:
\be
\label{qualresult3}
  \frac{\ccorr{ \gamma^2(\ve,T)}}{\gamma_{0}^2(\ve,T)}
  \sim
  \left( \frac{\EFGR}{\max\{ \ve, T\}} \right)^4 .
\ee
The temperature and the excitation energy enter the result in a similar way [note, however, the presence of the factor $\Upsilon(\eps/2T)$ in Eq.~(\ref{gamma-Upsilon}), that can change by two orders of magnitude]. This fact is a consequence of the lowest-order approximation. In higher orders their role is expected to be different, in accordance with the difference between the relaxation dynamics in the hot-electron and thermal problems \cite{Mirlin2015}.

It is important that in the range of applicability of the FGR description, $\max\{\eps,T\}\gg\EFGR$, the hybridiation with distant generations and eventually many-body localization effects (if any) become relevant at sufficiently large time scales. The characteristic time $t_*$ when $P_\text{FGR}(t)$ crosses over to a weaker dependence is determined by the initial state.
In the limit $\max\{\eps,T\}\gg\EFGR$, the scale $t_*$ satisfies $\gamma_0(\eps,T)t_*\gg1$, indicating that almost all the quasiparticle weight is lost during the FGR exponential relaxation
(for the hot-electron problem, $\gamma_0(\eps,0)t_* \sim \ln(\eps/\EFGR)$ \cite{Silvestrov2001}).

Our approach is conceptually similar to the one used by Basko, Aleiner and Altshuler (BAA)~\cite{BAA}, and hence it is instructive to compare the two results.
BAA considered inelastic relaxation in a chaotic quantum dot (in Ref.~\cite{BAA}, referred to as the localization cell), working in the basis of exact one-particle states $|\alpha\rangle$ and treating electron-electron interaction phenomenologically. Interaction matrix elements, $\corr{\alpha\beta|V|\gamma\delta}$, were assumed to be independent normally distributed random variables with the standard deviation $\lambda_\text{BAA}\Delta$ for energy difference smaller than the ultraviolet cutoff $M\Delta$, and zero otherwise.
The FGR relaxation rate is then given by $\gamma_{\text{BAA}}(T)\sim \lambda_\text{BAA}^2 MT$, and BAA obtained the following expression for  mesoscopic fluctuations:
\be
\label{mesofluct-BAA}
  \frac{\ccorr{\gamma^2(T)}_{\text{BAA}}}{\gamma^2_{\text{BAA}}(T)}
  \sim
  \frac{\lambda_\text{BAA}^4M\Delta^2T}{\gamma_{\text{BAA}}^3(T)}
  \sim
  \frac{\Delta^2}{\lambda_\text{BAA}^2M^2T^2}.
\ee
In order to apply these results to the case of a diffusive quantum dot, one should put $\lambda_\text{BAA} \sim \lambda \Delta/\ETh$~\cite{AGKL} and $M\sim T/\Delta$.
Then $\gamma_{\text{BAA}}(T)$ coincides with the Sivan-Imry-Aronov relaxation rate (\ref{SIAT}), while the estimate (\ref{mesofluct-BAA}) reproduces our result (\ref{qualresult3}) for mesoscopic fluctuations.

Finally, we emphasize that our result (\ref{subfinal}) for the leading contribution to mesoscopic fluctuations provides an exact account for the (screened) electron-electron interaction in a diffusive quantum dot. It has an important implementation for the statistics of the interaction matrix elements $\corr{\alpha\beta|V|\gamma\delta}$ in the lattice-model language. Since Ref.~\cite{AGKL}, it is usually assumed for simplicity that this statistics is Gaussian. However, our result (\ref{subfinal}) demonstrates that for a real diffusive quantum dot such an assumption is generally incorrect. Indeed, for the Gaussian statistics all correlators of the four matrix elements in $\ccorr{\gamma^2(\ve,T)}$ can be expressed through pairwise correlators (the Wick theorem), which are known to be determined by $E_2$ only \cite{Blanter,AGKL}.
Therefore the presence of the quantity $E_4$ in Eq.~(\ref{subfinal}) indicates that the statistics of the interaction matrix elements in a quantum dot is essentially non-Gaussian. This fact should be taken into account in constructing the theory of many-body localization in quantum dots.

We thank
I. Aleiner,
Ya.~M. Blanter,
M. V. Feigel'man,
I. V. Gornyi,
V. E. Kravtsov,
A. D. Mirlin,
and A. Silva
for useful discussions.
This work was supported by the Russian Science Foundation under Grant
No.\ 14-42-00044 (M.A.S.).

\appendix
\renewcommand*{\thefigure}{\arabic{figure}}

\section{Non-RPA averaging of the relaxation rate}
\label{S:Non-RPA}

In this Appendix
we consider the non-RPA diagrams \ref{F:2Coulomb-av}(b), \ref{F:2Coulomb-av}(c) and \ref{F:2Coulomb-av}(d) for the average energy relaxation rate.
The main difference from the RPA diagram \ref{F:2Coulomb-av}(a) is that now one [\ref{F:2Coulomb-av}(c) and \ref{F:2Coulomb-av}(d)] or both [\ref{F:2Coulomb-av}(b)] SSI lines
are coupled to diffusons via the square box
\be
\label{I-box}
  I
  =
  \,
  \raisebox{-8mm}{\includegraphics[width=17.7mm]{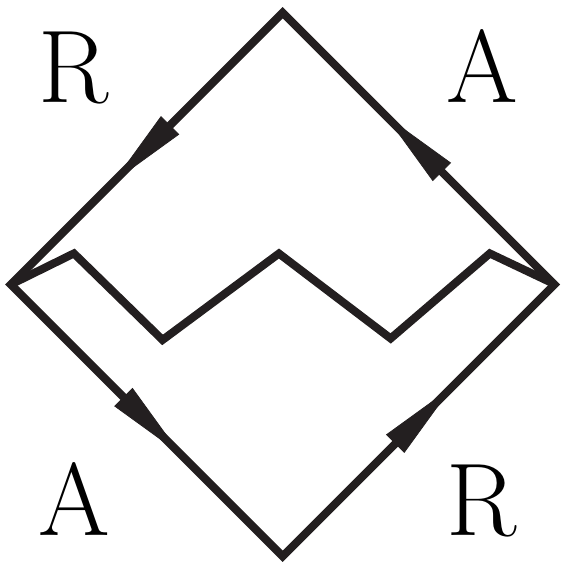}}
  \,
  =
  \int d\br \, [(G^\text{R}G^\text{A})(r)]^2 V(r) ,
\ee
where solid lines denote electron Green functions and the zigzag line stands for SSI.
Contrary to the RPA diagram \ref{F:2Coulomb-av}(a), where the momentum of the SSI line coincides with the slow ($q\sim1/L$) diffuson momentum, in the box configuration the interaction momentum can be comparable to $\pf$ typically carried by the Green functions.
The ratio of the diagram with the square box (\ref{I-box}) to a similar diagram with the RPA averaging is characterized by the parameter
\be
\label{f-I}
  \frs
  = \frac{I}{(2\pi\nu\tau)^2V(q=0)}
  = \frac 1{(2\pi\nu\tau)^2} \int d\br \, [(G^\text{R}G^\text{A})(r)]^2 \delta_{\kappa}(\br) .
\ee
where the smeared delta function $\delta_\kappa(\br)$ is defined in Eq.~(\ref{V(r)}).

The real-space representation of $(G^\text{R}G^\text{A})(r)$ in the limit of a good metal, $p_Fl\gg1$, is given by
\be
\label{GRGA}
  (G^\text{R}G^\text{A})(r)=\int \frac{e^{i\bp\br}}{\xi^2+1/4\tau^2} \frac{d\bp}{(2\pi)^d}
  = 2\pi\nu\tau \, j_d(\pf r) e^{-r/2l},
\ee
where $\tau$ and $l$ are the mean free time and mean free path correspondingly, $\xi = (p^2-\pf^2)/2m$, and the function
$j_d(x)=\corr{e^{ixn_1}}_\bn$ describes Friedel oscillations in $d$ dimensions:
$j_2(x) = J_0(x)$ and $j_3(x) = \sin x/x$ [$J_0(x)$ is the Bessel function].
Substituting Eq.~(\ref{GRGA}) into Eq.~(\ref{f-I}) we obtain for $\frs$:
\be
  \frs
  =
  \int j_d^2(\pf r) e^{-r/l} \delta_{\kappa}(\br) \, d\br
  =
  \int v_{\kappa}(\bq) \frac{d\bq}{(2\pi)^d}
  \int j_d^2(\pf r)e^{i\bq\br} e^{-r/l} \, d\br .
\ee
Then we average over angles and take the leading term in $\pf l\gg1$:
\be
  \frs
  =
  \int v_{\kappa}(\bq) \frac{d\bq}{(2\pi)^d}
  \int j_d^2(\pf r) j_d(qr) \, d\br .
\ee
Evaluating the integral over $\br$ we arrive at Eq.~(\ref{f-gen}).

In the case of the Coulomb interaction, $v_\kappa(\bq)$ is given by Eq.~(\ref{delta-kappa-Coul}) and we obtain \cite{AAL,Rosenbaum}
\be
\label{f-Coulomb}
  \frs_\text{3D}^\text{Coul}
  =
  \left( \frac{\kappa}{2\pf} \right)^2
  \ln \left[ 1 +  \left( \frac{2\pf}{\kappa} \right)^2 \right] ,
\qquad
  \frs_\text{2D}^\text{Coul}
  =
  \frac{2}{\pi}
  \frac{\mathop{\rm arccosh}(2\pf/\kappa)}{\sqrt{(2\pf/\kappa)^2-1}} .
\ee

\section{Expressions for $Y_{abcd}(\ve,\ve',\omega,\bq)$}
\label{A:Y}

Here we present explicit expressions for the effective interaction propagators
$Y_{abcd}(\ve,\ve',\omega,\bq)$ defined in Eq.~(\ref{Y-def}).
All components of $V=V(\omega,\bq)$ are taken at frequency $\omega$ and momentum $\bq$.
\begin{align}
&Y_{\RR,\RR}=V^\text{K}+\FF_{\ve'-\omega}V^\text{R}+\FF_{\ve}V^\text{A} , \\
&Y_{\RR,\RA}=(\FF_{\ve'}-\FF_{\ve'-\omega})(V^\text{K}+\FF_{\ve}V^\text{A})-(1-\FF_{\ve'}\FF_{\ve'-\omega})V^\text{R} \label{RRRA} , \\
&Y_{\RR,\AR}=-V^\text{R} , \\
&Y_{\RR,\AA}=V^\text{K}-\FF_{\ve'}V^\text{R}+\FF_{\ve}V^\text{A} , \\
&Y_{\RA,\RR}=-(\FF_{\ve}-\FF_{\ve-\omega})(V^\text{K}+\FF_{\ve'-\omega}V^\text{R})-(1-\FF_{\ve}\FF_{\ve-\omega})V^\text{A} , \\
&Y_{\RA,\RA}= -(\FF_{\ve}-\FF_{\ve-\omega})(\FF_{\ve'}-\FF_{\ve'-\omega})V^\text{K}+ (\FF_{\ve}-\FF_{\ve-\omega})(1-\FF_{\ve'}\FF_{\ve'-\omega})V^\text{R} \nonumber \\
&\hspace{15.5mm}{} -(1-\FF_{\ve}\FF_{\ve-\omega}) (\FF_{\ve'}-\FF_{\ve'-\omega})V^\text{A}  , \\
&Y_{\RA,\AR}=(\FF_{\ve}-\FF_{\ve-\omega})V^\text{R} , \\
&Y_{\RA,\AA}=-(\FF_{\ve}-\FF_{\ve-\omega})(V^\text{K}-\FF_{\ve'}V^\text{R})-(1-\FF_{\ve}\FF_{\ve-\omega})V^\text{A} \label{RAAA} , \\
&Y_{\AR,\RA}=-(\FF_{\ve'}-\FF_{\ve'-\omega})V^\text{A} , \\
&Y_{\AR,\AR}=0 , \\
&Y_{\AR,\AA}=-V^\text{A} , \\
&Y_{\AR,\RR}=-V^\text{A} , \\
&Y_{\AA,\RA}=(\FF_{\ve'}-\FF_{\ve'-\omega})(V^\text{K}-\FF_{\ve-\omega}V^\text{A})-(1-\FF_{\ve'}\FF_{\ve'-\omega})V^\text{R} , \\
&Y_{\AA,\AA}=V^\text{K}-\FF_{\ve'}V^\text{R}-\FF_{\ve-\omega}V^\text{A} , \\
&Y_{\AA,\RR}=V^\text{K}-\FF_{\ve-\omega}V^\text{A}+\FF_{\ve'-\omega}V^\text{R} , \\
&Y_{\AA,\AR}=-V^\text{R} .
\end{align}

\section{Reducible part of the correlator $\langle \gamma^2(\ve,T)\rangle$ \label{reducible part}}

In this methodological Appendix we show how to extract the reducible part of the correlator $\langle \gamma^2(\ve,T)\rangle$ from the general expression (\ref{corr}) and demonstrate that it coincides with the square $\gamma_0^2(\ve,T)$ of the average relaxation rate given by Eq. (\ref{SIAT}).
For simplicity, we consider the case of a sufficiently long-range interaction with $\frs=0$, when only the term $X_4X_4'$ contributes to Eq.~(\ref{corr}).

\begin{figure}
\centerline{\includegraphics[width=0.2\textwidth]{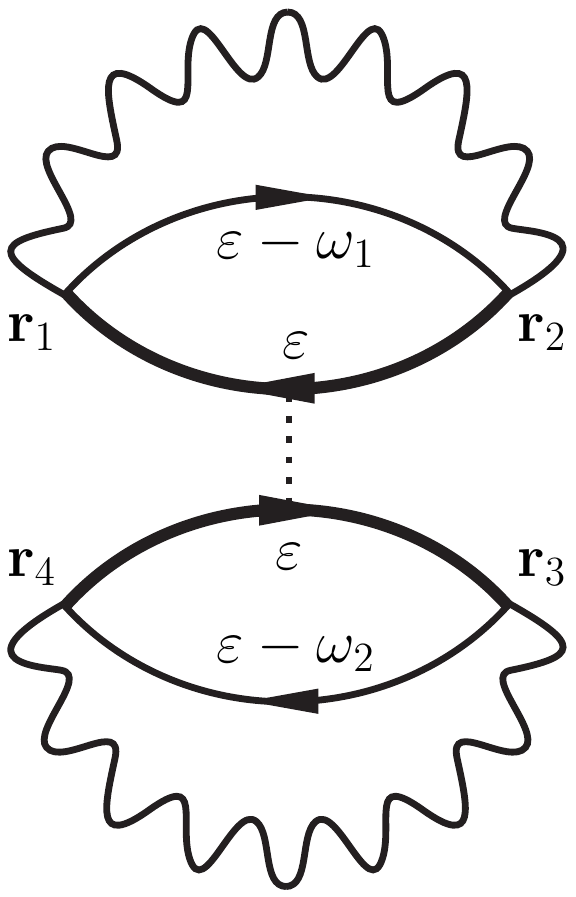}}
\caption{The reducible part of mesoscopic fluctuations of the relaxation rate, $\corr{\gamma^2(\ve,T)}_\text{red}$, in terms of the dynamically screened interaction [cf.\ Fig.~\ref{F:SIA2}(b)].
The dotted line denotes non-perturbative averaging of $G_\ve G_\ve$ over disorder.}
\label{SIASIA}
\end{figure}

In the formalism of kinetic equation used in our analysis, extraction of the reducible part should be done with care, and a naive independent averaging of $X_4$ and $X_4'$ would give a wrong answer.
The reason is that the product $\Delta G_\eps \Delta G_\eps$ contained both on the left- and right-hand sides of Eq.~(\ref{corr}) requires non-perturbative treatment.
Disorder averaging on the left-hand side of Eq.~(\ref{corr}) is discussed in Sec.~\ref{paircorr}, while on the right-hand side one has to take into account non-perturbative averaging between Green functions with the energies $\ve$ only.
In the language of the sigma model developed in Sec.~\ref{SS:SUSY}, this means that after averaging over fast modes non-perturbative integration over the sigma-model manifold should be performed only in the sector $1$ of the $1234$ space, whereas in all other sectors one can simply take the saddle-point value $Q_j=\Lambda$.

This observation significantly simplifies further calculation, as now we can use the representation of Eqs.~(\ref{gamma}) and (\ref{Sigma1}) in terms of the dynamically screened RPA interaction $\Delta V = V^\text{R} - V^\text{A} = 2i\Im V^\text{R}$ instead of that of Eq.~(\ref{corr}) with the statically screened interaction. The corresponding reducible contribution to $\langle \gamma^2(\ve,T)\rangle$ then should be extracted from
\begin{multline}
\langle \gamma^2(\ve,T)\rangle_{\text{red}}\int d\br \, d\br'\, \langle \Delta G_{\ve}(\br,\br) \Delta G_{\ve}(\br',\br') \rangle
\\{}
=\frac 14 \int d \br_1 \ldots d\br_4 (d\omega_1)(d\omega_2)\left(\BB_{\omega_1}+\FF_{\ve-\omega_1}\right)
\left(\BB_{\omega_2}+\FF_{\ve-\omega_2}\right)
\Delta V(\omega_1,\br_1-\br_2)\Delta V(\omega_2,\br_3-\br_4) \langle \tilde X_4 \rangle, \label{corSIA}
\end{multline}
where
\be
\tilde X_4 =  \Delta G_{\ve}(\br_2,\br_1)\Delta G_{\ve-\omega_1}(\br_1,\br_2)\Delta G_{\ve}(\br_4,\br_3)\Delta G_{\ve-\omega_2}(\br_3,\br_4),
\label{tilde X4}
\ee
and $\langle \ldots \rangle$ implies now averaging over minimal number (two) of fast modes and exact non-perturbative averaging of $G_{\ve}(\br_2,\br_1)G_{\ve}(\br_4,\br_3)$,
see the diagram in Fig.~\ref{SIASIA}.

Before we proceed, we highlight the important subtlety regarding R--A counting.
Consider first independent averaging over disorder in each bubble, when correlations between $G_{\ve}(\br_2,\br_1)$ and $G_{\ve}(\br_4,\br_3)$ are neglected.
This is precisely the situation encountered in Sec.~\ref{SS:SIA}, see Eq.~(\ref{<GG>}) and Fig.~\ref{F:SIA2}(c). To organize a diffuson in each bubble, one needs one $G^\text{R}$ and one $G^\text{A}$. Therefore only four out of sixteen terms would contribute to $\langle \Delta G_{\ve}(\br_2,\br_1)\Delta G_{\ve-\omega_1}(\br_1,\br_2)\rangle\langle\Delta G_{\ve}(\br_4,\br_3)\Delta G_{\ve-\omega_2}(\br_3,\br_4)\rangle$: RARA, RAAR, ARRA, ARAR.
The situation changes if we consider non-perturbative correlations between $G_{\ve}(\br_2,\br_1)$ and $G_{\ve}(\br_4,\br_3)$. In this case, only RAAR and ARRA from the previous four contributions survive. At the same time, now we have to take into account six additional terms: RRAR, RAAA, RRAA, ARRR, AARA, AARR. It turns out that each of this terms admits configuration with two fast diffusons, and, as a result, contributes equally to $\langle \gamma^2(\ve,T)\rangle_{\text{red}}$.

To perform averaging over disorder, we generalize the technique developed in Secs.~\ref{S:nonpert} and \ref{S:final}. The pair coupling introducing the 12 space is obvious: $\{G_{\ve}(\br_2,\br_1), G_{\ve}(\br_4,\br_3)\} $ and $\{  G_{\ve-\omega_1}(\br_1,\br_2), G_{\ve-\omega_2}(\br_3,\br_4) \}$. After perturbative averaging over fast modes and non-perturbative averaging over zero modes we obtain
\begin{gather}
\langle G^\text{R}G^\text{R}G^\text{A}G^\text{R} \rangle =-2(\pi\nu)^2X_1D(\br_1,\br_2)D(\br_3,\br_4), \label{GRRAR}
\\
\langle G^\text{R}G^\text{A}G^\text{A}G^\text{R} \rangle
  =
  (\pi\nu)^2
  \bigl[
  (4+2X_1+2X_2+4X_1X_2)D(\br_1,\br_2)D(\br_3,\br_4)
  + 4Z_1Z_2D(\br_1,\br_4)D(\br_2,\br_3)
  \bigr],
\\
\langle G^\text{R}G^\text{R}G^\text{A}G^\text{A} \rangle
  =
  (\pi\nu)^2
  \bigl[
  (4+2X_1+2X_2^*+4X_1X_2^*)D(\br_1,\br_2)D(\br_3,\br_4)
  + 4Z_1Z_2^*D(\br_1,\br_4)D(\br_2,\br_3)
  \bigr],
\\
\langle G^\text{R}G^\text{A}G^\text{A}G^\text{A} \rangle =-2(\pi\nu)^2X_1D(\br_1,\br_2)D(\br_3,\br_4).
\label{GRAAA}
\end{gather}
Here the order of Green functions is the same as in Eq.~(\ref{tilde X4}), and the factors $X_j$, $Z_j$ are defined in Eqs.~(\ref{XjZj})--(\ref{Omega1234}).
There are also four equations which are complex conjugate to Eqs.~(\ref{GRRAR})--(\ref{GRAAA}) (obtained by replacing $G^\text{R}\leftrightarrow G^\text{A}$).

As the energies in the second pair, \{$\eps-\omega_1$, $\eps-\omega_2$\}, are not supposed to coincide, the terms with $X_2$ and $Z_2$ describing correlations between $G_{\ve-\omega_1}(\br_1,\br_2)$ and $G_{\ve-\omega_2}(\br_3,\br_4)$ contain additional smallness of $\Delta/\max\{\ve,T\}$ and thus can be safely neglected.
Retaining only the terms with $X_1=\Delta/\pi\gamma(\eps)\gg1$, we obtain that each average (\ref{GRRAR})--(\ref{GRAAA}) as well as its complex conjugate, in the leading order, contributes $ \gamma_{0}^2(\ve,T)/8$ to $\langle \gamma^2(\ve,T) \rangle_{\text{red}}$. Hence their sum equals $\langle \gamma^2(\ve,T)\rangle_{\text{red}}=\gamma_{0}^2(\ve,T)$, reproducing the square of the average relaxation rate, as expected.

\section{Diffuson with the electron-electron interaction\label{eldiffuson}}

In Sec.~\ref{S:MF}, we introduced the inelastic single-particle level broadening in Eq.~(\ref{width}). This energy level width plays an important role as a regularizer of the otherwise divergent delta functions in the process of non-perturbative averaging over disorder [see, e.~g., Eq.~(\ref{XXX})].
As has been pointed out in Ref.~\cite{Lopes}, the same quantity $\gamma_0(\ve,T)$
determines the width of the single-particle electron Green function (that was calculated in Sec.~\ref{S:Average}) and the mass of two-electron particle-particle (cooperon) \cite{FukuyamaAbrahams,Fukuyama} and particle-hole (diffuson) \cite{Blanter,Lopes,Castellani86} propagators in the presence of interaction. To make this paper comprehensive, we rederive this result in the framework of the modified Keldysh technique introduced in Sec.~\ref{SS:Keldysh}. We consider the zero-momentum diffuson and assume Coulomb interaction in the limit $\frs\to0$.

\begin{figure}
\centerline{\includegraphics[width=.7\textwidth]{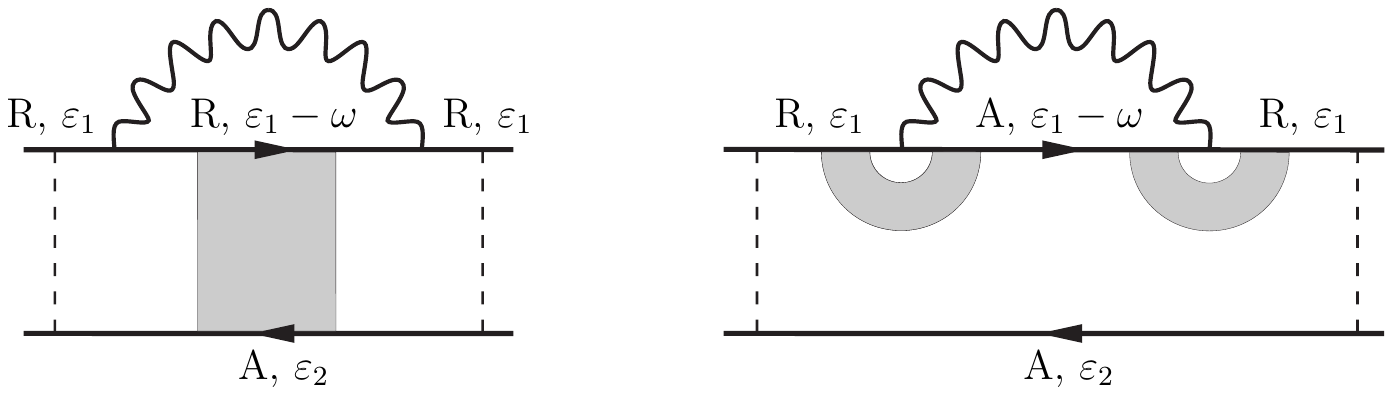}}
\caption{
Diagrams for the diffuson self-energy $\Pi(\eps_1,\eps_2)$, with the interaction line inserted into $G^\text{R}$ (there are mirror counterparts with the interaction inserted into $G^\text{A}$).
The bare diffuson $D_0(\omega, \bq)$ is depicted by the shaded block.
The dashed lines denote the beginning of external diffusons which are not included into the self-energy part.
The diagram (b) admits insertion of an impurity line in two ways (not shown), producing the standard Hikami box as an interaction vertex of four diffusons \cite{Hikami}.}
\label{diffuson3}
\end{figure}

The important contribution comes from the elastic processes only (processes that do not lead to the energy exchange between $G^\text{R}_{\ve_1}$ and $G^\text{A}_{\ve_2}$). As a result, the diffuson $D_{\text{el}}(\ve_1,\ve_2)$ can be found from the algebraic (not integral) Dyson equation:
\be
D^\text{R}_{\text{el}}(\ve_1,\ve_2)=D^\text{R}_0(\ve_1-\ve_2,0)+D^\text{R}_0(\ve_1-\ve_2,0)\Pi(\ve_1,\ve_2) D^\text{R}_{\text{el}}(\ve_1,\ve_2).
\ee
Here $D^\text{R}_0(\omega,\bq)$ is the bare diffuson, defined in Eq.~(\ref{<GG>}), and $\Pi(\ve_1,\ve_2)$ is the elastic part of the self-energy (also at zero momentum). In the lowest order, $\Pi(\ve_1,\ve_2)$ is given by the diagrams shown in Fig.~\ref{diffuson3} and their counterparts where the interaction line is coupled to $G^\text{A}$.

The contribution of the diagrams in Fig.~\ref{diffuson3} equals
\begin{align}
  \Pi_1(\ve_1,\ve_2)
  =
  &-\frac i2 \int (d\omega) (d\bq) D^\text{R}_0(\ve_1-\ve_2-\omega,\bq)\left(V^\text{K}+\FF_{\ve_1-\omega}V^\text{R}+\FF_{\ve_1}V^\text{A}\right)
\nonumber
\\{}
  &+\frac i2\int (d\omega)(d\bq)
  \left[Dq^2-i(\ve_1-\ve_2+\omega)\right]
  \left(D^\text{R}_0(\omega, \bq)\right)^2\left(\FF_{\ve_1}-\FF_{\ve_1-\omega}\right)V^\text{R} .
\label{Pi1}
\end{align}
The contribution of the analogous diagrams with the interaction line inserted to $G^A$ can be written as
\begin{align}
  \Pi_2(\ve_1,\ve_2)
  = &-\frac i2 \int (d\omega) (d\bq) D^\text{R}_0(\ve_1-\ve_2-\omega,\bq)\left(V^\text{K}-\FF_{\ve_2}V^\text{A}-\FF_{\ve_2+\omega}V^\text{R}\right)
\nonumber
\\{}
  & +\frac i2\int (d\omega)(d\bq)
  \left[Dq^2-i(\ve_1-\ve_2+\omega)\right]
  \left(D^\text{R}_0(\omega, \bq)\right)^2\left(\FF_{\ve_2+\omega}-\FF_{\ve_2}\right)V^\text{R}.
\label{Pi2}
\end{align}
In Eqs.~(\ref{Pi1}) and (\ref{Pi2}) all interaction propagators are taken at frequency $\omega$ and momentum $\bq$.
Employing analyticity properties, we obtain for the diffuson self-energy $\Pi=\Pi_1+\Pi_2$:
\begin{multline}
\label{Pi3}
  \Pi(\ve_1,\ve_2)
  =
  -\frac i2 \int (d\omega)(d\bq) D^\text{R}_0(\omega,\bq)
  \Bigl\{
    2V^\text{K}(\ve_1-\ve_2-\omega,\bq)
\\{}
  +(\FF_{\omega-\ve_1}+\FF_{\omega+\ve_2})\left[V^\text{R}(\ve_1-\ve_2-\omega,\bq)-V^\text{R}(\omega,\bq)+i(\ve_1-\ve_2)D^\text{R}_0(\omega,\bq)V^\text{R}(\omega,\bq) \right]
  \Bigr\}.
\end{multline}

In the universal limit of Coulomb interaction, $V^{\text{R}(\text{A})}(\omega,\bq)=(Dq^2\mp i\omega)/\ns\nu Dq^2$, we obtain
\be
  \Pi(\ve_1,\ve_2)
  =
  -\frac 1{\ns\nu}\int \frac{(d\omega)(d\bq)\omega(2\BB_{\omega}-(\FF_{\omega-\ve_2}+\FF_{\omega+\ve_1}))}{Dq^2(Dq^2-i\omega-i(\ve_1-\ve_2))}
  +\frac{\ve_1-\ve_2}{\ns\nu}\int \frac{(d\omega)(d\bq)(\FF_{\omega-\ve_2}+\FF_{\omega+\ve_1})}{Dq^2(Dq^2-i\omega-i(\ve_1-\ve_2))}.
\ee
Two terms in this expression have a different nature. In the first term, the integration over $\omega$ converges at the scale $\max\{T,\ve_1,\ve_2\}$, and the summation over momenta converges at $Dq^2\sim \ETh$. In the second term, the main contribution comes from $\ETh\ll Dq^2 \ll \omega \ll \tau^{-1}$ ($\tau$ is the mean free time). Assuming the equilibrium distribution function, we obtain
\be
\Pi(\ve_1,\ve_2)=-\frac{\Delta}{2\pi \ns E_2^2}\left(2\pi^2 T^2 +\ve_1^2+\ve_2^2\right)
+\frac {2(\ve_1-\ve_2)}{\ns\nu}\int\frac{(d\omega) (d\bq) \sign \omega}{Dq^2 (Dq^2-i\omega)}.
\ee
Using the result (\ref{SIAT}) for $\gamma_{0}(\ve,T)$ (with $\lambda=1$ and $\frs=0$), we can write
\be
  \Pi(\ve_1,\ve_2)=-\frac{\gamma_{0}(\ve_1,T)+\gamma_{0}(\ve_2,T)}2 +i(\ve_1-\ve_2)(Z-1),
\ee
where $Z$ is the wave-function renormalization factor \cite{Castellani84}:
\be
Z-1=\frac 2{\pi \ns\nu} \int_{1/L}^{1/l}\frac{(d\bq)}{Dq^2} \ln\frac 1{Dq^2\tau},
\ee
$L$ is the linear size of the sample, $l$ is the mean free path.

Finally, the zero-dimensional diffuson acquires the form
\be
  D^\text{R}_{\text{el}}(\ve_1,\ve_2)
  =
  \frac 1{-i(\ve_1-\ve_2)Z+[\gamma_{0}(\ve_1,T)+\gamma_{0}(\ve_2,T)]/2} .
\ee
Precisely the same form of the diffuson follows from the sigma model (\ref{action}), provided the wave function renormalization can be neglected, $Z-1\ll 1$.

\section{Contraction rules for the zero-dimensional sigma model}
\label{slowcontrrules}

In Efetov's parametrization \cite{Efetov-book}, the sigma-model action for $\corr{G^\text{R}_{\ve_1}G^\text{A}_{\ve_2}}$
[where the Green functions may contain a finite width $\gamma(E)$, see Eq.~(\ref{width})] depends only on the `radial' variables $\lambda_F$ and $\lambda_B$:
\be
\label{S0-a}
S_{0}=a\left(\lambda_{\B}-\lambda_{\F}\right),
\ee
where
\be
a=-\frac{\pi i}{\Delta}\left\{(\ve_1)_+ - (\ve_2)_-\right\},
\ee
and $\ve_{\pm}$ are defined in Eq. (\ref{width}).
In order to calculate correlation functions, the full form of the parametrization should be employed.
After some superalgebra, one can verify that the following non-perturbative contraction rules hold for averaging over the zero-dimensional sigma model with the action (\ref{S0-a}):
\begin{subequations}
\label{contr-rules-0D}
\begin{gather}
\label{contr-rules-0D-1}
\langle \str PQ\str RQ\rangle=\str P\Lambda\str R\Lambda +Z \str\left(PR-P\Lambda R\Lambda\right)-X\left(\str P \str R -\str P\Lambda \str R\Lambda\right),
\\
\label{contr-rules-0D-2}
\langle\str PQ RQ\rangle=\str P\Lambda R\Lambda +Z\left(\str P\str R- \str P\Lambda \str R\Lambda\right)-X\str\left(PR-P\Lambda R\Lambda\right),
\end{gather}
\end{subequations}
where $P$ and $Q$ are arbitrary supermatrices, and the functions $X(a)$ and $Z(a)$ are defined as
\begin{gather}
X(a)=\frac 12\int_0^1 d\lambda_{\F}\int _1^{\infty} d\lambda_{\B} e^{-a(\lambda_{\B}-\lambda_{\F})}=\frac{1-e^{-2a}}{2a^2},
\label{X}
\\
Z(a)=\frac 12\int_0^1 d\lambda_{\F}\int _1^{\infty} d\lambda_{\B}\frac{\lambda_{\B}+\lambda_{\F}}{\lambda_{\B}-\lambda_{\F}}e^{-a(\lambda_{\B}-\lambda_{\F})}= \frac 1a.
\label{Z}
\end{gather}

\section{Spurious pairings of eight Green functions}
\label{A:spurious}

\begin{figure}
\centerline{\includegraphics[width=0.9\textwidth]{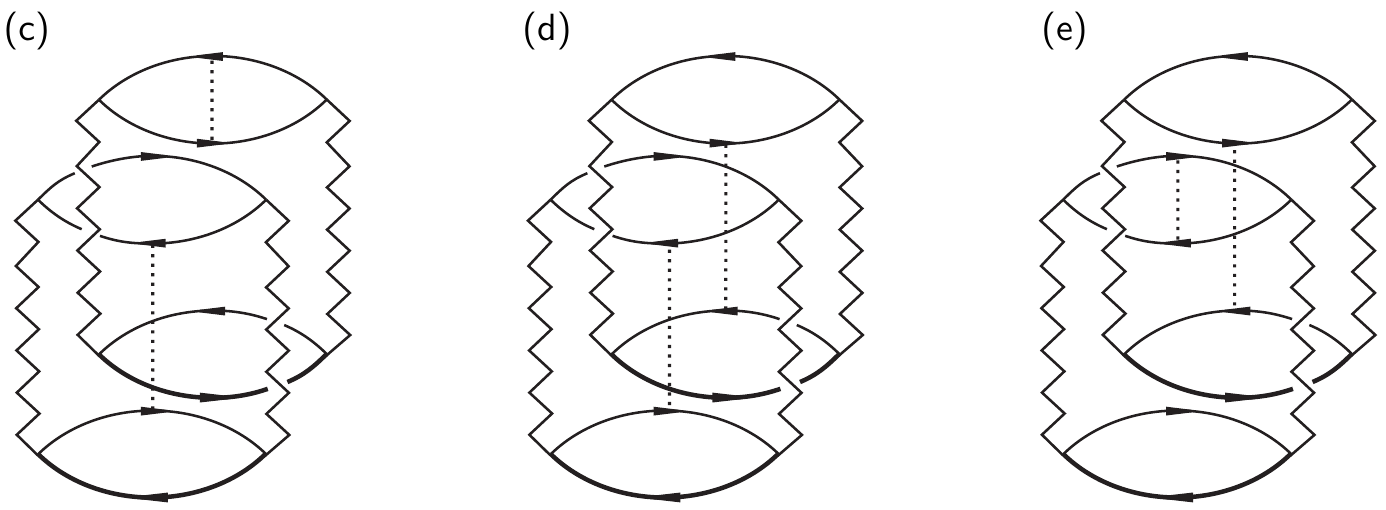}}
\caption{
Spurious ways to match the energies of eight Green functions listed in Eq.~(\ref{energy-matching-spurious}). All energy and coordinate indices are the same as in Fig. \ref{F:AGKL31}(a). Two pairs are shown by the dotted lines, while the rest four Green functions have the energy $\eps$ of the decaying particle. Such configurations automatically nullify the collision integral and do not contribute to mesoscopic fluctuations of $\gamma(\eps,T)$.}
\label{F:spurious}
\end{figure}

As we explained in Sec.\ \ref{slowaveraging}, the maximal number of the zero-argument delta-functions $\delta_{\gamma}(0)$ in the expression for $\corr{\gamma^2(\eps,T)}$ originates if two Green functions with energy $\ve$ are paired together, and the remaining six Green functions are grouped such that energy coincidence in two pairs automatically guarantees the energy coincidence in the third pair.
Direct analysis shows, however, that apart from two physical possibilities (a) and (b), given by  Eq.~(\ref{energy-matching}) and shown in Fig.~\ref{F:AGKL31}, there are three others that can formally satisfy this condition:
\be
\label{energy-matching-spurious}
  \text{(c) \:  $\ve_1 \approx \ve$ and $\omega_2 \approx 0$},
\qquad
  \text{(d) \: $\ve_1 \approx \ve_2 \approx \ve$} ,
\qquad
  \text{(e) \: $\ve_2 \approx \ve$ and $\omega_1 \approx 0$}.
\ee
However, these new possibilities shown in Fig.~\ref{F:spurious} appear to be too restrictive: In each of these diagrams, the energies of four Green functions are pinned to the external energy $\ve$, while the rest four are grouped in two pairs (shown by the dotted line). The possibilities listed in Eq.~(\ref{energy-matching-spurious}) are unphysical as they do not describe any real decay process (the final state consists of a single electron instead of an electron and an electron-hole pair) and one can expect that they do not contribute to the relaxation rate. Indeed, using expressions from \ref{A:Y} one can immediately check that spurious contributions with either $\ve_1=\ve$ or $\ve_2=\ve$ nullify the collision integral. Consequently, the only possibility to acquire an additional large factor $\delta_{\gamma}(0)$ is to analyze two pairings (a) and (b), which is performed in the main text.

%\section*{References}

\end{document}